\documentclass[%
 reprint,
superscriptaddress,
 amsmath,amssymb,
 prapplied,
aps,
]{revtex4-2}

\usepackage{graphicx}
\usepackage{dcolumn}
\usepackage{bm}
\usepackage{braket}
\usepackage{hyperref}
\hypersetup{
    colorlinks=true,
    linkcolor=blue,     
    urlcolor=blue,
    citecolor=blue,
}
\usepackage[capitalise]{cleveref}
\usepackage[euler]{textgreek}
\usepackage{xcolor}
\usepackage[version=4]{mhchem}
\usepackage{physics}
\usepackage{float}

\begin{document}


\title{Studying Critical Parameters of Superconductor via Diamond Quantum Sensors}

\author{Kin On Ho}
 \thanks{These authors contributed equally to this work.}
\affiliation{
Department of Physics, The Chinese University of Hong Kong, Shatin, New Territories, Hong Kong, China
}
\affiliation{
Department of Physics, The Hong Kong University of Science and Technology, Clear Water Bay, Kowloon, Hong Kong, China
}
\author{Wai Kuen Leung}
 \thanks{These authors contributed equally to this work.}
\affiliation{
Department of Physics, The Hong Kong University of Science and Technology, Clear Water Bay, Kowloon, Hong Kong, China
}
\author{Yiu Yung Pang}
 \thanks{These authors contributed equally to this work.}
\affiliation{
Department of Physics, The Chinese University of Hong Kong, Shatin, New Territories, Hong Kong, China
}
\author{King Yau Yip}
\author{Jianyu Xie}
\affiliation{
Department of Physics, The Chinese University of Hong Kong, Shatin, New Territories, Hong Kong, China
}
\author{Yi Man Liu}
\author{Aliki Sofia Rotelli}
\author{Man Yin Leung}
\author{Ho Yin Chow}
\affiliation{
Department of Physics, The Hong Kong University of Science and Technology, Clear Water Bay, Kowloon, Hong Kong, China
}
\author{Kwing To Lai}
\affiliation{
Department of Physics, The Chinese University of Hong Kong, Shatin, New Territories, Hong Kong, China
}
\author{Andrej Denisenko}
\affiliation{Max Planck Institute for Solid State Research, Stuttgart, Germany}
\affiliation{3. Physikalisches Institut, Integrated Quantum Science and Technology (IQST), University of Stuttgart, Pfaffenwaldring 57, 70569 Stuttgart, Germany}
\author{B. Keimer}
\affiliation{Max Planck Institute for Solid State Research, Stuttgart, Germany}
\author{J\"{o}rg Wrachtrup}
\affiliation{Max Planck Institute for Solid State Research, Stuttgart, Germany}
\affiliation{3. Physikalisches Institut, Integrated Quantum Science and Technology (IQST), University of Stuttgart, Pfaffenwaldring 57, 70569 Stuttgart, Germany}
\author{Sen Yang}
 \email{phsyang@ust.hk}
\affiliation{
Department of Physics, The Chinese University of Hong Kong, Shatin, New Territories, Hong Kong, China
}
\affiliation{
Department of Physics, The Hong Kong University of Science and Technology, Clear Water Bay, Kowloon, Hong Kong, China
}

\date{\today}

\begin{abstract}
Critical parameters are the key to superconductivity research, and reliable instrumentations can facilitate the study. Traditionally, one has to use several different measurement techniques to measure critical parameters separately. In this work, we develop the use of a single species of quantum sensor to determine and estimate several critical parameters with the help of independent simulation data. We utilize the nitrogen-vacancy (NV) center in the diamond, which recently emerged as a promising candidate for probing exotic features in condensed matter physics. The non-invasive and highly stable nature provides extraordinary opportunities to solve scientific problems in various systems. Using a high-quality single-crystalline \ce{YBa_{2}Cu_{4}O_{8}} (YBCO) as a platform, we demonstrate the use of diamond particles and a bulk diamond to probe the Meissner effect. The evolution of the vector magnetic field, the $H-T$ phase diagram, and the map of fluorescence contour are studied via NV sensing. Our results reveal different critical parameters, including lower critical field $H_{c1}$, upper critical field $H_{c2}$, and critical current density $j_{c}$, as well as verifying the unconventional nature of this high-temperature superconductor YBCO. Therefore, NV-based quantum sensing techniques have huge potential in condensed matter research.
\end{abstract}

\maketitle


\section{Introduction}
Condensed matter physics is one of the most significant fields in modern physics. With improved knowledge of material physics, many scientific problems have been solved in the last century. Measurement techniques such as resistivity, superconducting quantum interference device (SQUID), and neutron scattering, are state-of-the-art and powerful in modern research. Future research demands a highly consistent tool for studying critical phenomena. Despite the huge success classical sensors offer in studying sample properties, they are sometimes limited in capturing certain exotic features in modern research such as magnetic properties under high pressure. Moreover, methods without contact or gauging are robust since they do not damage the treasured sample.

\begin{figure}[t]
\includegraphics[width=8.6cm]{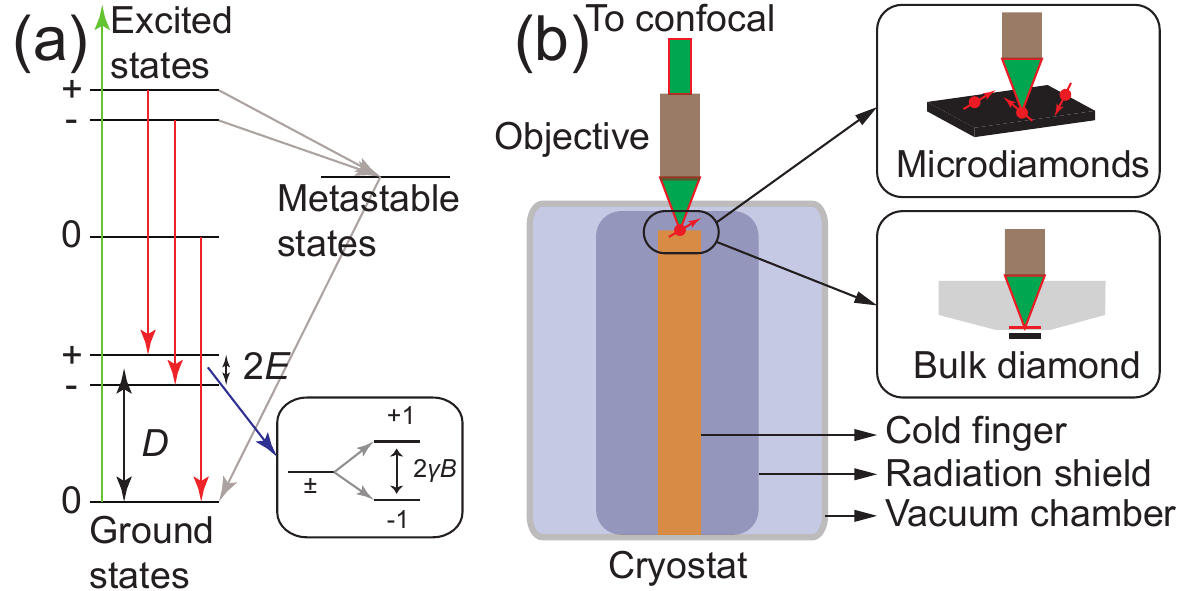}
\caption{(a) The spin-state dependent transition structures. The $\ket{m_{s} = \pm 1}$ states emit less fluorescence than $\ket{m_{s} = 0}$ state due to the inter-system crossing to metastable states. (b) The schematic apparatus of our experiments. A home-built confocal microscope is aligned with a Montana cryostat. Either MDs (top) or bulk diamonds (bottom) are used for different sensing purposes.}
\label{fig1}
\end{figure}

Quantum sensors are proposed to be a solution to these problems. Recently, negatively charged nitrogen-vacancy (NV) centers in diamonds have emerged as an extraordinary candidate for condensed matter physics research \cite{Bouchard2011Detection, Nusran2018Spatially, Joshi2019Measuring, Xu2019Mapping, Thiel2016Quantative, Pelliccione2016Scanned, Schlussel2018Wide, Ho2021Diamond, McLaughlin2021Strong, Monge2023Spin, Casola2018Probing, Lillie2020Laser, Rohner2018Real, Scheidegger2022Scanning, Paone2021All, Joshi2020Quantum}, especially as they retain their robustness under extreme conditions, such as low temperatures and high pressures \cite{Yip2019Measuring, Lesik2019Magnetic, Hsieh2019Imaging, Ho2020Probing, Ho2021Recent, Toraille2020Combined, Ho2023Spectroscopic, Hilberer2023Enabling}. The NV center is formed by a vacancy in the carbon lattice adjacent to a nitrogen atom, which gives a high spatial resolution. The negatively charged state is a spin-1 system that has exceptional sensitivity, especially for magnetic fields. The simplified energy structures are shown in \cref{fig1}(a), depicting the spin-state dependent fluorescence of $\ket{m_{s}= 0}$ and $\ket{m_{s}= \pm 1}$ sublevels. Hence, the optical initialization and readout can be achieved simply by using a green laser. Whenever the microwave (MW) frequency is in resonance with the ground-state sublevel transitions ($\ket{m_{s} = 0} \rightarrow \ket{m_{s} = \pm 1}$), the NV center emits less fluorescence due to the inter-system crossing to the metastable states. By analyzing the fluorescence against MW frequencies, the information encoded in the ground state can be extracted. This method is the well-known technique called optically detected magnetic resonance (ODMR) \cite{Gruber1997Scanning}.

Attributed to the diverse nature of independent instrumentations, usually only a single physical quantity can be probed with one technique. Hence, several experiments are implemented in studying one sample. Given the robustness of NV centers in capturing magnetic signals, independent works measure separately either the critical temperatures and/or critical fields \cite{Bouchard2011Detection, Yip2019Measuring, Lesik2019Magnetic, Hsieh2019Imaging, Joshi2019Measuring, Joshi2020Quantum, Nusran2018Spatially} or critical currents \cite{Paone2021All}. Motivated by previous works, we focus on employing only NV centers in determining several critical parameters of the Meissner effect with minimum requirements on instrumentation. The Meissner effect, which is the signature of a superconductor, is the repulsion of a surrounding magnetic field. Thus, we keep track of the magnetic field profile during the superconducting phase transition. In this work, in contrast to analyzing only the ODMR splitting, we first discuss the framework of the magnetic field which is more intuitive and informative in superconductivity. We reveal that the ODMR method is powerful enough to extract lower and upper critical temperatures $T_{c1}$ and $T_{c2}$ without knowing the exact magnetic fields. With these understandings, we show that the use of a bulk diamond is very versatile. The edge contour method, which is benchmarked with the ODMR method, not only can be used to extract $T_{c1}$ and $T_{c2}$ but also verify independent simulation data that is used to estimate the critical current density $j_{c}$. The measurements on the bulk diamond are performed using a standard confocal setup, minimizing the cost of instruments and sample preparation.

Another focus of this work is the high-temperature cuprate superconductor, which is an important candidate to study unconventional superconductivity. Compared to \ce{YBa_{2}Cu_{3}O_{7-$\delta$}}, \ce{YBa_{2}Cu_{4}O_{8}} (YBCO) is comparatively less studied \cite{Schilling1990Specific, Bucher1990Anisotropic, Varshney1996Coherence, Panagopoulos1999Effects, Basov1994Caxis, Basov1995InPlane, Grissonnanche2014Direct, Heyen1991Two, Khasanov2023Oxygen, Khasanov2008S}. YBCO has one of the largest tuning potentials to study cuprate superconductors, thus it is meaningful to develop techniques, in addition to traditional methods, to study this precious material. A high-quality and single-crystalline unconventional high-temperature superconductor YBCO is chosen for the demonstration. Here, we utilize microdiamonds (MDs) with an average size of 1 \textmu m and an anvil-shaped bulk diamond in close proximity to investigate the temperature-dependent stray magnetic fields surrounding the sample. We reconstruct the magnetic field profile experienced by the MDs using ODMR spectroscopy. We also map the fluorescence contour using a bulk diamond. Characteristic parameters, such as lower critical field $H_{c1}$ and upper critical field $H_{c2}$, can be determined from our data. The critical current density $j_{c}$ is also estimated based on the fluorescence contour mapping with the help of simulation data. Hence showcasing the excellent potential of the NV center in probing exotic features for condensed matter research.

\section{Method}
\subsection{Experimental details}
For the experiments using MDs, we first prepare a solution of 1-\textmu m MDs with a 3-ppm NV concentration. The solution is ultra-sounded for 10 minutes and subsequently a direct dropcast onto the YBCO surface is performed. During the evaporation of the solution, we gently move the sample using a needle in order to have MDs on both sample surfaces. After the solution is dried, we transfer the sample to the experimental setup carefully using a needle.

For the experiments using an anvil-type bulk diamond, we first spin coat a thin layer of polymethyl methacrylate (PMMA) on the implanted cutlet of the anvil. The thickness of the PMMA is estimated to be roughly 1 \textmu m. We then stick the YBCO onto the PMMA layer. A non-magnetic metal structure is made to fix the anvil on the side without blocking the light path. Hence, the diamond is inverted so that the laser and fluorescence can pass through the anvil from the back side.

To perform optical measurements at low temperatures, our home-built confocal microscope \cite{Ho2020Probing} is aligned with a Montana cryostat. A long working-distance Mitutoyo Plan Apo SL objective (100X for MDs; 50X for bulk diamond) is placed outside the cryostat for better fluorescence collection. A micro-coil is used as a waveguide for a uniform MW transmission \cite{Ho2020Probing}.

\subsection{NV Physics}
To understand the ODMR spectroscopy, we consider the following effective ground-state Hamiltonian written in the $\ket{m_{z}}$ basis:
\begin{equation}
H_{g}^{eff} = D S_{z}^{2} + E (S_{x}^{2} - S_{y}^{2}) + \gamma_{e} \vec{B} \cdot \vec{S},
\label{Hgeff}
\end{equation}
where $S$ is the spin-1 operator, $D$ and $E$ are the longitudinal and transverse zero-field splitting (ZFS) parameters respectively, $\gamma_{e} = 2.8$ MHz/G is the gyromagnetic ratio and $B$ represents the external magnetic field. $D$ is mainly sensitive to the temperature and pressure \cite{Doherty2014Temperature, Chen2011Temperature, Acosta2010Temperature, Doherty2014Electronic, Ho2020Probing, Steele2017Optically}, while $E$ indicates the strain or local charge environment \cite{Barson2017Nanomechanical, Mittiga2018Imaging, Ho2020Probing, Ho2021In}. In our data analysis, the temperature-dependent $D$ term is carefully calibrated using the known empirical data \cite{Chen2011Temperature}.

With the presence of longitudinal and transverse magnetic fields ($B_{\parallel}$ and $B_{\perp}$), the four NV axes have different responses in the ODMR spectrum \cite{Shin2013Suppression, Doherty2014Measuring}. Here, we adopt the second-order perturbation theory \cite{Doherty2014Measuring} and the magnetic field can be expressed as
\begin{align}
\label{Bperp} B_{\perp} &= \frac{1}{\sqrt{3}\gamma_{e}} \sqrt{D \left( f_{+} + f_{-} \right) -2D^2},\\
\label{Bparallel} |B_{\parallel}| &= \frac{1}{2\gamma_{e} D} \sqrt{D^2 \left( f_{+} - f_{-} \right)^2 - \gamma_{e}^4 B_{\perp}^4},\\
\label{Btot} B_{tot} & = \sqrt{B_{\perp}^{2}+B_{\parallel}^{2}},\\
\label{Btheta} B_{\theta} &= \arctan{\left( \frac{B_{\perp}}{\abs{B_{\parallel}}} \right)},
\end{align}
where the upper frequency $f_{+}$ is the resonance-on-the-right and the lower frequency $f_{-}$ is the resonance-on-the-left of the same NV axis, respectively. These equations allow one to calculate the magnetic field $B_{\parallel}$ and $B_{\perp}$ in the local frame of the NV center. With proper assumptions and analysis, we derive the change in magnetic field vectors across the superconducting phase transition using a two-cone method (see the vector approach in \cite{SI}).

\section{Results}
\subsection{Constructing the YBCO phase diagram via MD}
\begin{figure}[t]
\includegraphics[width=8.6cm]{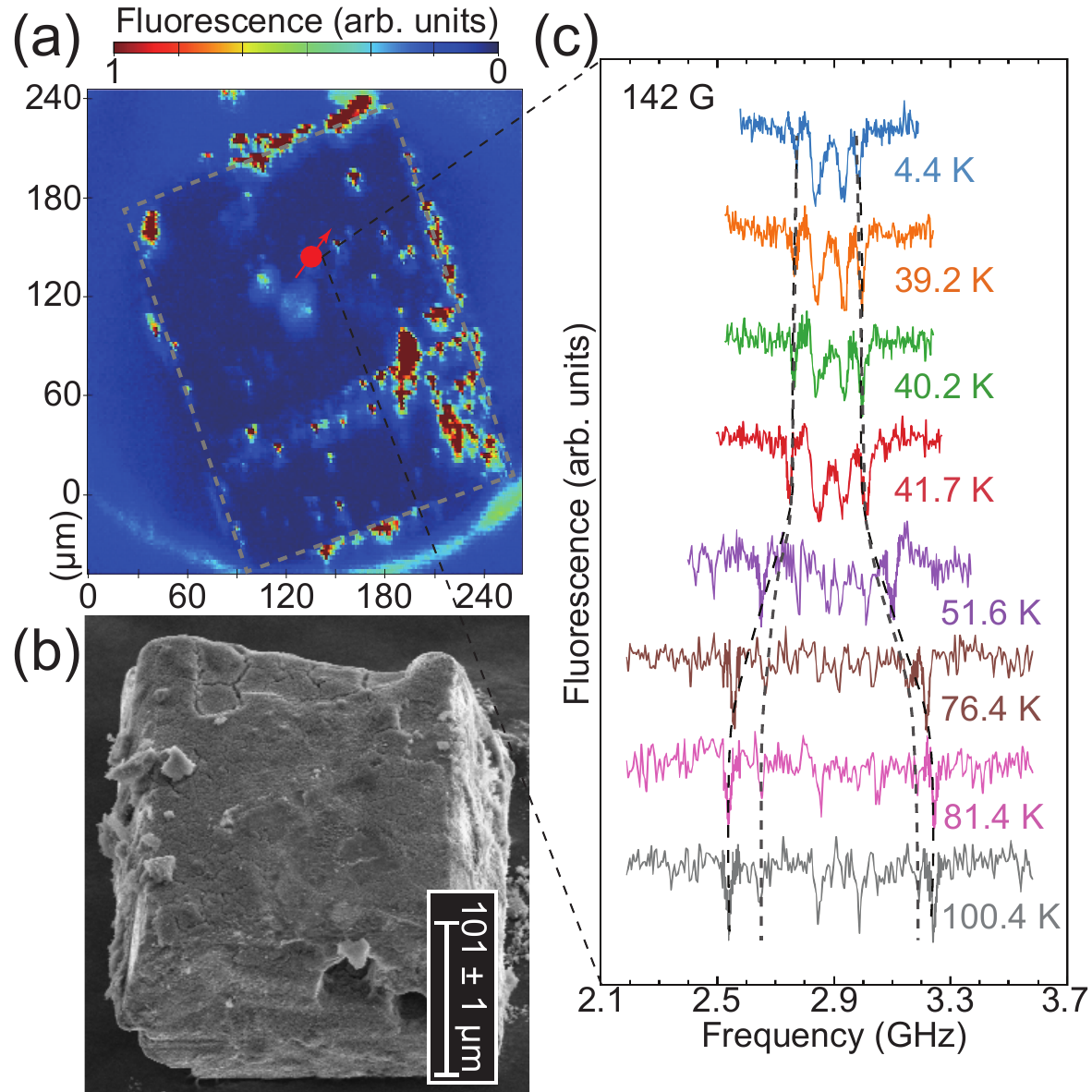}
\caption{(a) A confocal image of a YBCO surface with MDs dropcasted on top of it. One MD close to the center of the sample was chosen for the subsequent measurements. (b) A SEM image shows the thickness of the sample. The dimensions determined from the SEM measurement are $239 \pm 1$ (length) x $162 \pm 1$ (width) x $101 \pm 2$ (thickness) \textmu m$^{3}$. (c) Selected ODMR trace of the chosen MD at different temperatures under 142 G. Due to the Meissner effect of the YBCO, the magnetic field sensed by the MD varies with temperature, resulting in changes in the ODMR splitting. The black-dash line tracks the outermost splitting, while the grey-dash line tracks the second outermost splitting against temperatures.}
\label{fig2}
\end{figure}

The confocal image in \cref{fig2}(a) shows the YBCO and the locations of the MDs, determined by the bright spots. One single-crystalline MD close to the center of the sample is chosen for the subsequent measurements. Since confocal microscopy cannot determine the sample thickness, scanning electron microscope (SEM) images are taken to determine the sample dimensions as shown in \cref{fig2}(b). \cref{fig2}(c) shows the selected ODMR trace of the chosen MD at different temperatures under an external magnetic field of 142 G (determined by MD at normal states). The ODMR splitting has a sharp change with temperature, indicating changes in the YBCO sample's effective demagnetization due to the Meissner effect. These changes directly reflect the superconducting state of the YBCO sample. Hence, the splitting against temperature data can be extracted to analyze the superconducting phase transition (frequency approach \cite{SI}).

\begin{figure}[t]
\includegraphics[width=8.6cm]{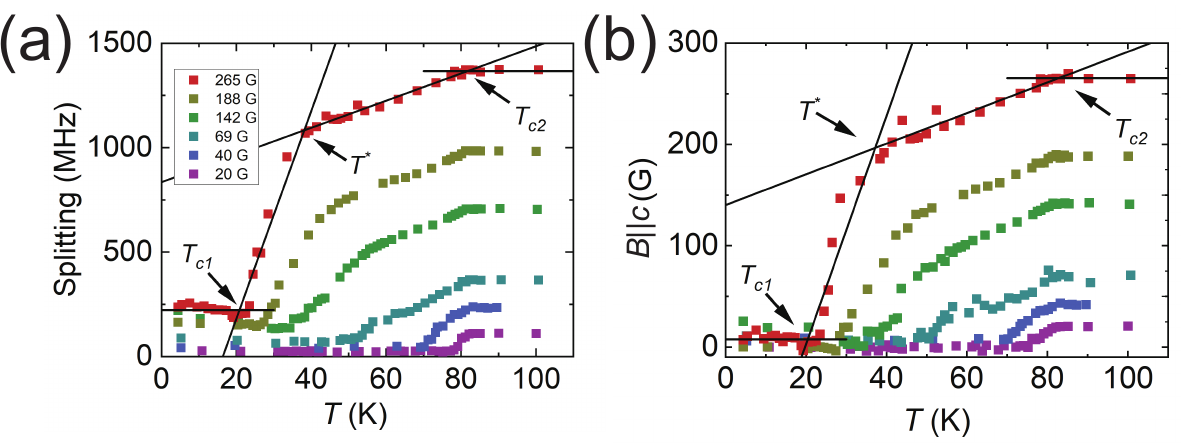}
\caption{(a) The ODMR splitting against temperature under different applied magnetic fields. (b) The derived magnetic field $B_{\parallel c}$ against temperature under different applied magnetic fields. In general, both splitting and $B_{\parallel c}$ have similar behaviors. Three turning points ($T_{c1}, T^{*}, T_{c2}$) can be observed during warming up the system. The applied magnetic fields are calculated from the ODMR spectrum at normal states.}
\label{fig3}
\end{figure}

To determine the characteristic temperatures, we employ a four-line fitting approach similar to Ref. \cite{Krusin-Elbaum1989Temperature} assuming there are no sharp features during the superconducting phase transition (i.e. no kinks or jumps). The splitting and the magnetic field against temperatures are fitted using four straight lines. In the data processing, two baselines (Meissner and normal states) are set to be horizontal and the best fit for the appropriate four straight lines is performed. The $T_{c1}$, $T^{*}$, and $T_{c2}$ are then determined from the interceptions of the lines (see \cref{fig3}). $T_{c1}$ and $T_{c2}$ refer to the transition from the Meissner to vortex state and from the vortex to normal state, respectively. $T^{*}$ refers to a slope change in the vortex state and further discussion can be found in \cite{SI, Bean1962Magnetization, Bean1964Magnetization, Krusin-Elbaum1989Temperature, Krusin-Elbaum1990New, Tomioka1994The, Wolf1993Relaxation, Kumar1989Extension, Brandt1993TypeII, Brandt1996Geometric, Abrikosov1957The, Thiel2016Quantative, Pelliccione2016Scanned, Schlussel2018Wide, Auslaender2009Mechanics, Lillie2020Laser}. Moreover, the $T_{c2}(B = 20 \text{ G})$ obtained at the lowest applied field is treated as the critical temperature $T_{c} = 81.8 \pm 1.6$ K for the subsequent normalization. We check this by the AC susceptibility measurement in another setup and find $T_{c}^{\text{AC}} = 78.0$ K, which is in good agreement with our ODMR method.

We conduct measurements similar to \cref{fig2}(c) at different applied magnetic fields, and the ODMR splittings are extracted and plotted in \cref{fig3}(a). In general, we observe no changes in $T_{c2}$, and systematic changes in $T_{c1}$ and $T^{*}$ with respect to the applied magnetic field. These splittings are further converted to the stray magnetic field along the sample $c$-axis (\cref{fig3}(b)) using the two-cone method discussed in \cite{SI}.

\begin{figure}[t]
\includegraphics[width=8.6cm]{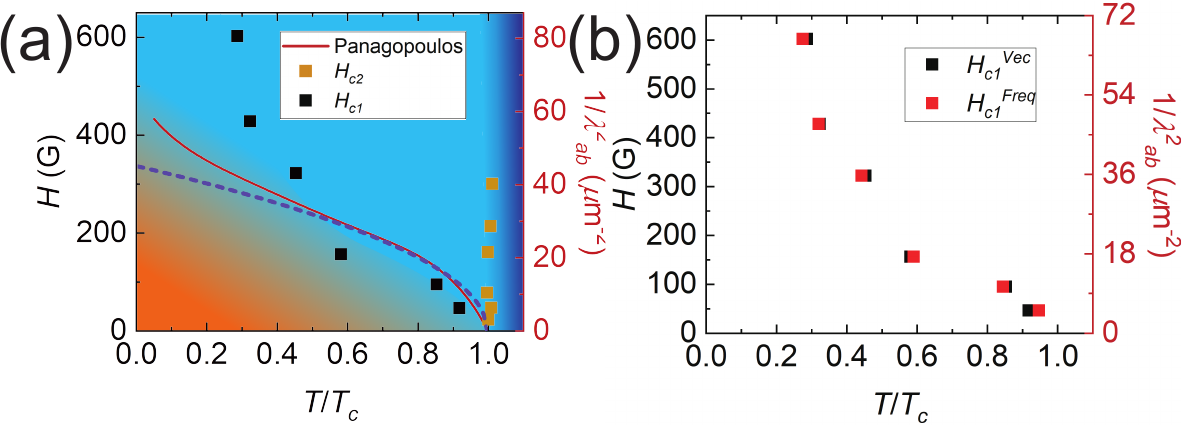}
\caption{(a) The $H-T$ phase diagram of the YBCO. These values are determined from the vector approach in \cref{fig3}(b). The violet-dash line represents a conventional BCS theory, with $H_{c1}(0) = 339$ G \cite{Varshney1996Coherence}, which goes as $\sqrt{(1-T/T_{c})}$ and our $H_{c1}$ data show evidence of a violation. We also have a steeper turn compared to Ref. \cite{Panagopoulos1999Effects}. The color in the background is a guide to the eye. (b) The comparison between frequency and vector approaches in constructing the $H-T$ phase diagram. The $H_{c1}$ are in excellent agreement except for the low field region, since the vector field is more sensitive in the magnetic field direction compared to the splitting change.}
\label{fig4}
\end{figure}

Finally, the $H-T$ phase diagram of YBCO is constructed following Ref. \cite{Prozorov2018Effective, SI}. We first discuss the phase diagram done by the vector approach. As shown in \cref{fig4}(a), the $H_{c1}$ data show evidence of previously claimed $s$+$d$-wave superconductor, which is obviously violating the traditional BCS theory which goes as $\sqrt{(1-T/T_{c})}$. Our data is comparable with the previous results \cite{Panagopoulos1999Effects}, but we have a steeper turn in $H_{c1}$ presumably due to differences in bulk and powder samples. The $H_{c2}$ data is almost unchanged as expected. Since $H_{c2}$ is related to the effective mass of the electron, it has negligible change with respect to a small magnetic field. Then, we compare the phase diagram extracted from the vector and frequency approaches. As shown in \cref{fig4}(b), two data sets perfectly match each other except for the lowest field data point. The discrepancy can be explained by the mixing of the response from four NV axes, and the fact that the vector field is more sensitive compared to the splitting change since $B_{\perp}$ shifts but not splits the resonances. The estimated error bound is around 0.02 $T/T_{c}$ considering the experimental data uncertainty and analysis resolution, which is enclosed with the size of the markers. Overall, the frequency approach is also valuable for determining the critical fields even if losing the information on the magnetic field profile.

\subsection{Mapping the fluorescence contour using a bulk diamond}
In contrast to the traditional wide-field imaging technique which maps the detailed magnetic field distribution using a camera, we aim to get a map of fluorescence contour using the same confocal setup (\cref{fig1}(b)) for the same YBCO. We target the fluorescence contour induced by the edge stray field of the YBCO, hence we first search for the region with the highest magnetic field by performing ODMR across the sample edge. Then we track this region in the subsequent temperature-dependent measurements to get the evolution of the fluorescence contrast of this contour (see Ref. \cite{SI}). Note that the contrast derived in our scheme is, in principle, only affected by the region with the same targeted magnetic field. Hence, the region outside and inside the sample may have the same fluorescence (contrast) but different magnetic field profiles and only the fluorescence along the edge contour shows a change in contrast.
\begin{figure}[t]
\includegraphics[width=8.6cm]{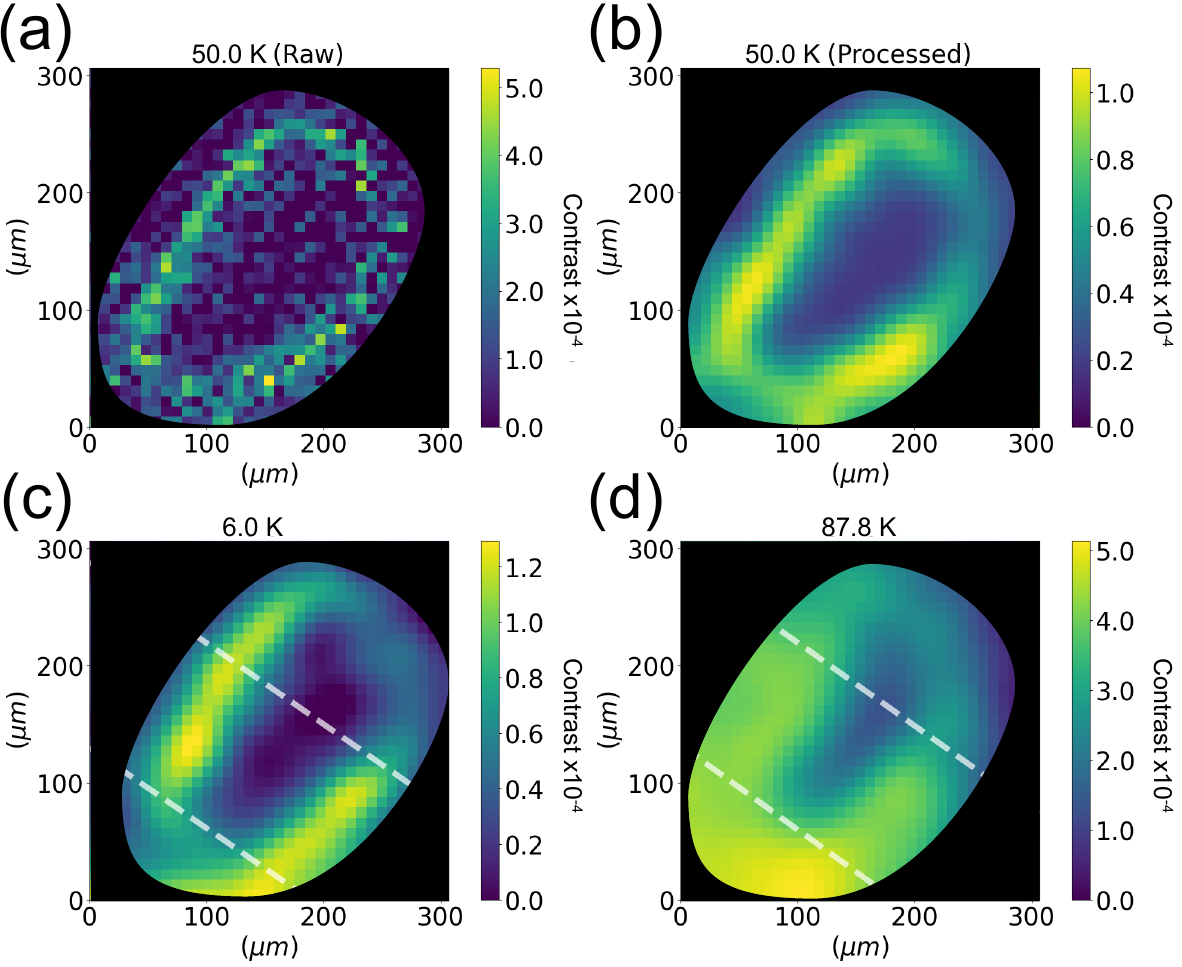}
\caption{(a) The raw fluorescence contour image with $B_{\parallel c} \approx 34$ G at 50.0 K. The yellow color indicates the fluorescence contour with $B_{NV} \approx 43$ G along the sample edge. The difference in fluorescence is presented as the contrast. (b) Same as (a) but Gaussian filtering is used to improve the sharpness for visualization. See text for details. (c) The fluorescence contour image at 6.0 K ($B_{NV} \approx 49$ G). (d) The fluorescence contour image at 87.8 K ($B_{NV} = B_{\parallel c} \approx 34$ G). The fluorescence contour flattens towards the center of the sample upon warming up the system. The black color masks the region far away from the sample for better visualization.}
\label{fig5}
\end{figure}

The applied magnetic field is $B_{\parallel c} \approx 34$ G. A raw fluorescence contour image with $B_{NV} \approx 43$ G at 50.0 K is shown in \cref{fig5}(a), and the fluorescence contour can be observed along the sample edge. The enhancement in the magnetic field is from the Meissner-induced edge current in shielding the external magnetic field. To improve the sharpness for visualization, we employ Gaussian filtering \cite{Gao1993A, Knikker2004A, Itatani2013Intraventricular, Capellera2021Heat} and the result is shown in \cref{fig5}(b). The filter is applied to average out noises and fluctuations within a local area on the sample during the measurement, enhancing the signal-to-noise ratio of our image for further analysis of the phase transition. The fluorescence contour images are recorded during warming up the system. The fluorescence contour images at 6.0 K ($B_{NV} \approx 49$ G) and 87.8 K ($B_{NV} = B_{\parallel c} \approx 34$ G) are shown in \cref{fig5}(c, d), revealing a significant change of the fluorescence contour towards the center of the sample. The darker color inside the sample even at the normal state (\cref{fig5}(d)) is from an intrinsically lower contrast region (about a half), which is probably due to the imperfectness of the implantation and MW transmission.

\begin{figure}[t]
\includegraphics[width=8.6cm]{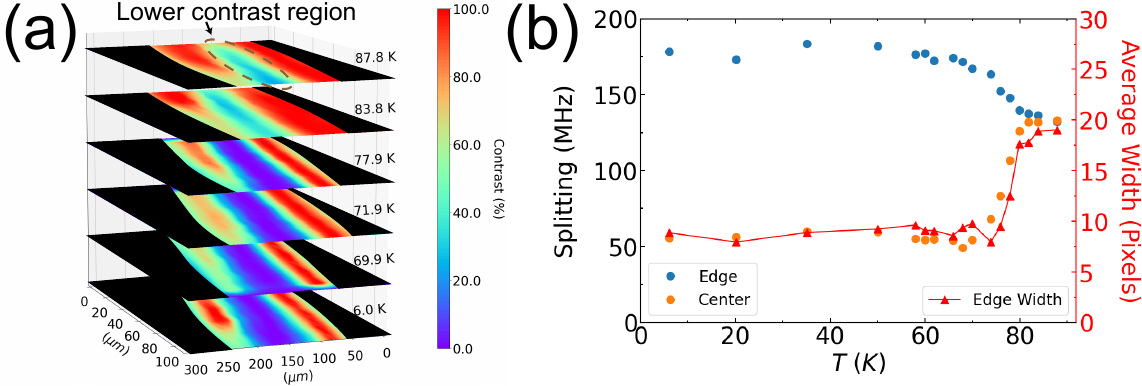}
\caption{(a) The 2D plot of the evolution of the edge fluorescence contour. The contour gradually flattens towards the center of the sample while warming up. This evolution hints at a superconducting phase transition. For better visualization, the black color masks the region far away from the sample and the color scale represents the normalized contrast. (b) The splitting of the edge and center ODMR spectra (left-$y$) and the average pixel width of the contour image (right-$y$) against temperature. The contour width shows a similar behavior as the center ODMR splitting, indicating that it is possible to extract important parameters with the transition.}
\label{fig6}
\end{figure}

To get a qualitative analysis of the temperature evolution, we extract the data inside two white-line cuts along the width of the sample as shown in \cref{fig5}(c, d). We plot in \cref{fig6}(a) the 2D plot of the evolution of the map of fluorescence contour. Below $\approx 70$ K, no systematic change is observed, thus only the lowest temperature data is presented. While warming up the system above $\approx 70$ K, the contour gradually flattens towards the center of the sample. Moreover, the contour's width can be further used to determine the superconducting phase transition. The lines shown in \cref{fig5}(c, d) are further used to calculate the number of pixels of the contour. The criteria for extracting the width is set to be the full-width-half-maximum (FWHM). For a fair comparison with the MD measurement, we also measure the ODMR at the center of the sample. The results are presented in \cref{fig6}(b), revealing some transition-like features. The center ODMR splitting can be used to determine the critical temperatures with the same procedure as MD measurements, yet the edge ODMR splitting is not sensitive enough to determine those. Furthermore, the edge contour width shows a similar behavior as the center ODMR splitting, meaning that it is possible to extract critical temperatures $T_{c1}$ and $T_{c2}$ with the transition. The $T_{c1}$ and $T_{c2}$ determined from the edge contour width are {$72.3 \pm 1.4$ K} and $83.8 \pm 1.7$ K, while they are $71.0 \pm 1.4$ K and $81.7 \pm 1.6$ K determined from center ODMR splitting. Presumably, the 2-K step of taking data introduces a noticeable discrepancy when extracting the parameters. Yet they are still in good agreement.

\begin{figure}[t]
\includegraphics[width=8.6cm]{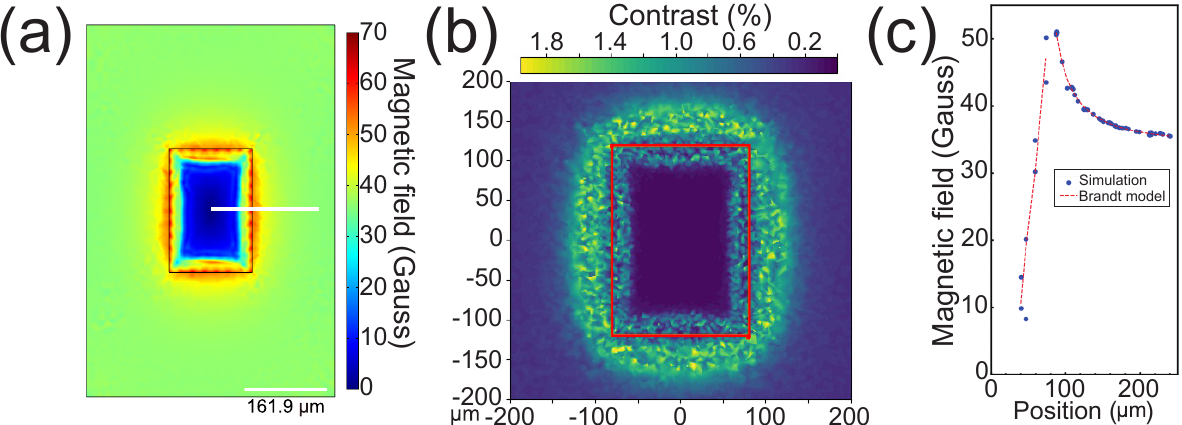}
\caption{(a) The 2D plot of the simulated stray magnetic field on the surface of the sample. The color scale shows the magnitude of the stray magnetic field. (b) The derived contrast plot on the edge contour. The red rectangle represents the sample geometry. (c) The fitting of Brandt's model to the white line cut of the magnetic field profile in (a).}
\label{fig7}
\end{figure}

To simulate the Meissner shielding of the YBCO at low temperatures, we perform finite-element analysis to solve the time-dependent Ginzburg-Landau (TDGL) equations \cite{Gulian2020Exploring}. This approach provides a straightforward verification of the map of fluorescence contours measured experimentally. We model our sample as a perfect rectangular cuboid with the dimensions 239.4 (length) x 161.9 (width) x 101.1 (thickness) \textmu m$^{3}$. The zero-temperature-limit London penetration depth $\lambda$ and coherence length $\xi$ are taken as 160.4 and 1.1 nm, respectively \cite{Varshney1996Coherence}. The applied magnetic field is set to be 34 G. We also set a long evolution time for reaching the equilibrium state. In \cref{fig7}(a), we present the simulated magnetic field of the sample's surface. We also find an edge fluorescence contour of $\approx 50$ G with a NV-YBCO distance of about 1 \textmu m, which is expected for our spin-coated PMMA layer. This fluorescence contour is in agreement with our measurement. However, we notice that the simulated center magnetic field (a few Gauss) is slightly lower than what we measured ($\approx 10$ G), presumably due to the imperfectness of the diamagnetism of the real YBCO sample. In \cref{fig7}(b), we convert the magnetic field mapping to the fluorescence contour plot by assuming a constant contrast and linewidth with respect to the magnetic field vector. Here, the experimental contrast and linewidth in the ODMR spectrum are taken as the reference. Although this overestimates the contour contrast and width, the edge contour is comparable with \cref{fig5}. Note that the scatter-like feature around the edge is due to the finite mesh size in the simulation software.

Furthermore, the magnetic field profile in \cref{fig7}(a) matches Brandt's model \cite{Brandt1993TypeII}, as shown in \cref{fig7}(c). Brandt's model suggests that

\begin{align}
H(y) &= \frac{J_{c}}{\pi} \arctan{\frac{\sqrt{(y^{2}-b^{2})}}{c\abs{y}}} \text{ for } y<a,\\
H(y) &= \frac{J_{c}}{\pi} \arctan{\frac{c\abs{y}}{\sqrt{(y^{2}-b^{2})}}} \text{ for } y>a,
\end{align}
where the sample's width $2a = 161.9$ \textmu m, $J_{c}$ is the critical sheet current and the fitting parameters $b$ and $c$ are defined as

\begin{align}
b &= a/\cosh{\frac{\pi H_{a}}{J_{c}}},\\
c &= \tanh{\frac{\pi H_{a}}{J_{c}}}.
\end{align}
The critical current density $j_{c} = J_{c}/d$, where $d = 101.1$ \textmu m is the sample thickness. The corresponding fitting parameters are $j_{c} = 5372$ A/cm$^{2}$, $b = 36.8$ \textmu m, and $c = 0.890$ for the field inside the sample, while $j_{c} = 8484$ A/cm$^{2}$, $b = 47.3$ \textmu m, and $c = 0.760$ for the field outside the sample. The $j_{c}$ value is slightly lower than some literature values ($\sim 10^{4}$ A/cm$^{2}$) \cite{Naito2016Critical, Lombardo2011Critical}, although some works show orders of magnitudes higher ($\sim 10^{6} - 10^{7}$ A/cm$^{2}$) \cite{Braccini2011Properties, Polak2002Magnetic, Beck2004Determination}. The discrepancies can be explained by (1) the different oxygen contents in the YBCO samples, which contribute to different critical properties, (2) the different geometries (filament against thin film against bulk) of the samples, and (3) the meshing of the simulation components. We suspect that the meshing of the system will be crucial for accurately capturing the value since we find that the boundary layer parameters will noticeably affect the simulation result. Optimizing these parameters requires, however, demanding computational resources. In general, this simulation provides a fair estimation of the sample characteristic parameters.

\section{Conclusion}
In this work, we showcase the capability of magnetic field sensing using NV centers for condensed matter physics research. We show that one can implement only the NV center with minimum requirements on the instrumentation to probe several critical parameters by analyzing the magnetic profile of the sample. The Meissner effect of the YBCO is probed via MDs and a bulk diamond which reveals the temperature-dependent magnetic responses. The critical fields ($H_{c1}$ and $H_{c2}$) and critical temperatures $T_{c}$ are extracted from the MDs. The critical current density $j_{c}$ is estimated from the simulation data, which is independently verified by the map of fluorescence contour imaged by the bulk diamond. Future work can implement a geometrical approach \cite{Maertz2010Vector, Zhang2022Single} and pulsed measurements \cite{Hsieh2019Imaging, McLaughlin2021Strong, Monge2023Spin, Tetienne2013Spin, Schmid-Lorch2015Relaxometry, Barry2020Sensitivity} with a proper design of the experimental apparatus. The excellent potential of NV centers in condensed matter research is proven important for developing NV-based sensing techniques.

\begin{acknowledgments}
We thank W. Zhang and Y.F. Chen for the technical support. We thank R. Lortz, H. C. Po, and S. K. Goh for the fruitful discussion. K.O.H acknowledges financial support from the Hong Kong PhD Fellowship Scheme. S.Y. acknowledges financial support from Hong Kong RGC (GRF/14304419). J.W. acknowledges funding by the DFG via GRK2642 and FOR 2724, the Max Planck Society and the BMBF via the cluster4future QSENS.

Kin On Ho, Wai Kuen Leung, and Yiu Yung Pang contributed equally to this work.
\end{acknowledgments}

\bibliography{references}

\begin{thebibliography}{73}%
\makeatletter
\providecommand \@ifxundefined [1]{%
 \@ifx{#1\undefined}
}%
\providecommand \@ifnum [1]{%
 \ifnum #1\expandafter \@firstoftwo
 \else \expandafter \@secondoftwo
 \fi
}%
\providecommand \@ifx [1]{%
 \ifx #1\expandafter \@firstoftwo
 \else \expandafter \@secondoftwo
 \fi
}%
\providecommand \natexlab [1]{#1}%
\providecommand \enquote  [1]{``#1''}%
\providecommand \bibnamefont  [1]{#1}%
\providecommand \bibfnamefont [1]{#1}%
\providecommand \citenamefont [1]{#1}%
\providecommand \href@noop [0]{\@secondoftwo}%
\providecommand \href [0]{\begingroup \@sanitize@url \@href}%
\providecommand \@href[1]{\@@startlink{#1}\@@href}%
\providecommand \@@href[1]{\endgroup#1\@@endlink}%
\providecommand \@sanitize@url [0]{\catcode `\\12\catcode `\$12\catcode `\&12\catcode `\#12\catcode `\^12\catcode `\_12\catcode `\%12\relax}%
\providecommand \@@startlink[1]{}%
\providecommand \@@endlink[0]{}%
\providecommand \url  [0]{\begingroup\@sanitize@url \@url }%
\providecommand \@url [1]{\endgroup\@href {#1}{\urlprefix }}%
\providecommand \urlprefix  [0]{URL }%
\providecommand \Eprint [0]{\href }%
\providecommand \doibase [0]{https://doi.org/}%
\providecommand \selectlanguage [0]{\@gobble}%
\providecommand \bibinfo  [0]{\@secondoftwo}%
\providecommand \bibfield  [0]{\@secondoftwo}%
\providecommand \translation [1]{[#1]}%
\providecommand \BibitemOpen [0]{}%
\providecommand \bibitemStop [0]{}%
\providecommand \bibitemNoStop [0]{.\EOS\space}%
\providecommand \EOS [0]{\spacefactor3000\relax}%
\providecommand \BibitemShut  [1]{\csname bibitem#1\endcsname}%
\let\auto@bib@innerbib\@empty
\bibitem [{\citenamefont {Bouchard}\ \emph {et~al.}(2011)\citenamefont {Bouchard}, \citenamefont {Acosta}, \citenamefont {Bauch},\ and\ \citenamefont {Budker}}]{Bouchard2011Detection}%
  \BibitemOpen
  \bibfield  {author} {\bibinfo {author} {\bibfnamefont {L.-S.}\ \bibnamefont {Bouchard}}, \bibinfo {author} {\bibfnamefont {V.~M.}\ \bibnamefont {Acosta}}, \bibinfo {author} {\bibfnamefont {E.}~\bibnamefont {Bauch}},\ and\ \bibinfo {author} {\bibfnamefont {D.}~\bibnamefont {Budker}},\ }\bibfield  {title} {\bibinfo {title} {Detection of the meissner effect with a diamond magnetometer},\ }\href {https://doi.org/10.1088/1367-2630/13/2/025017} {\bibfield  {journal} {\bibinfo  {journal} {New Journal of Physics}\ }\textbf {\bibinfo {volume} {13}},\ \bibinfo {pages} {025017} (\bibinfo {year} {2011})}\BibitemShut {NoStop}%
\bibitem [{\citenamefont {Nusran}\ \emph {et~al.}(2018)\citenamefont {Nusran}, \citenamefont {Joshi}, \citenamefont {Cho}, \citenamefont {Tanatar}, \citenamefont {Meier}, \citenamefont {Bud'ko}, \citenamefont {Canfield}, \citenamefont {Liu}, \citenamefont {Lograsso},\ and\ \citenamefont {Prozorov}}]{Nusran2018Spatially}%
  \BibitemOpen
  \bibfield  {author} {\bibinfo {author} {\bibfnamefont {N.~M.}\ \bibnamefont {Nusran}}, \bibinfo {author} {\bibfnamefont {K.~R.}\ \bibnamefont {Joshi}}, \bibinfo {author} {\bibfnamefont {K.}~\bibnamefont {Cho}}, \bibinfo {author} {\bibfnamefont {M.~A.}\ \bibnamefont {Tanatar}}, \bibinfo {author} {\bibfnamefont {W.~R.}\ \bibnamefont {Meier}}, \bibinfo {author} {\bibfnamefont {S.~L.}\ \bibnamefont {Bud'ko}}, \bibinfo {author} {\bibfnamefont {P.~C.}\ \bibnamefont {Canfield}}, \bibinfo {author} {\bibfnamefont {Y.}~\bibnamefont {Liu}}, \bibinfo {author} {\bibfnamefont {T.~A.}\ \bibnamefont {Lograsso}},\ and\ \bibinfo {author} {\bibfnamefont {R.}~\bibnamefont {Prozorov}},\ }\bibfield  {title} {\bibinfo {title} {Spatially-resolved study of the meissner effect in superconductors using {NV}-centers-in-diamond optical magnetometry},\ }\href {https://doi.org/10.1088/1367-2630/aab47c} {\bibfield  {journal} {\bibinfo  {journal} {New Journal of Physics}\ }\textbf {\bibinfo {volume} {20}},\ \bibinfo {pages} {043010}
  (\bibinfo {year} {2018})}\BibitemShut {NoStop}%
\bibitem [{\citenamefont {Joshi}\ \emph {et~al.}(2019)\citenamefont {Joshi}, \citenamefont {Nusran}, \citenamefont {Tanatar}, \citenamefont {Cho}, \citenamefont {Meier}, \citenamefont {Bud'ko}, \citenamefont {Canfield},\ and\ \citenamefont {Prozorov}}]{Joshi2019Measuring}%
  \BibitemOpen
  \bibfield  {author} {\bibinfo {author} {\bibfnamefont {K.}~\bibnamefont {Joshi}}, \bibinfo {author} {\bibfnamefont {N.}~\bibnamefont {Nusran}}, \bibinfo {author} {\bibfnamefont {M.}~\bibnamefont {Tanatar}}, \bibinfo {author} {\bibfnamefont {K.}~\bibnamefont {Cho}}, \bibinfo {author} {\bibfnamefont {W.}~\bibnamefont {Meier}}, \bibinfo {author} {\bibfnamefont {S.}~\bibnamefont {Bud'ko}}, \bibinfo {author} {\bibfnamefont {P.}~\bibnamefont {Canfield}},\ and\ \bibinfo {author} {\bibfnamefont {R.}~\bibnamefont {Prozorov}},\ }\bibfield  {title} {\bibinfo {title} {Measuring the lower critical field of superconductors using nitrogen-vacancy centers in diamond optical magnetometry},\ }\href {https://doi.org/10.1103/PhysRevApplied.11.014035} {\bibfield  {journal} {\bibinfo  {journal} {Phys. Rev. Applied}\ }\textbf {\bibinfo {volume} {11}},\ \bibinfo {pages} {014035} (\bibinfo {year} {2019})}\BibitemShut {NoStop}%
\bibitem [{\citenamefont {Xu}\ \emph {et~al.}(2019)\citenamefont {Xu}, \citenamefont {Yu}, \citenamefont {Hui}, \citenamefont {Su}, \citenamefont {Cheng}, \citenamefont {Chang}, \citenamefont {Zhang}, \citenamefont {Shen},\ and\ \citenamefont {Tian}}]{Xu2019Mapping}%
  \BibitemOpen
  \bibfield  {author} {\bibinfo {author} {\bibfnamefont {Y.}~\bibnamefont {Xu}}, \bibinfo {author} {\bibfnamefont {Y.}~\bibnamefont {Yu}}, \bibinfo {author} {\bibfnamefont {Y.~Y.}\ \bibnamefont {Hui}}, \bibinfo {author} {\bibfnamefont {Y.}~\bibnamefont {Su}}, \bibinfo {author} {\bibfnamefont {J.}~\bibnamefont {Cheng}}, \bibinfo {author} {\bibfnamefont {H.-C.}\ \bibnamefont {Chang}}, \bibinfo {author} {\bibfnamefont {Y.}~\bibnamefont {Zhang}}, \bibinfo {author} {\bibfnamefont {Y.~R.}\ \bibnamefont {Shen}},\ and\ \bibinfo {author} {\bibfnamefont {C.}~\bibnamefont {Tian}},\ }\bibfield  {title} {\bibinfo {title} {Mapping dynamical magnetic responses of ultrathin micron-size superconducting films using nitrogen-vacancy centers in diamond},\ }\href {https://doi.org/10.1021/acs.nanolett.9b02298} {\bibfield  {journal} {\bibinfo  {journal} {Nano Letters}\ }\textbf {\bibinfo {volume} {19}},\ \bibinfo {pages} {5697} (\bibinfo {year} {2019})},\ \bibinfo {note} {pMID: 31348663},\ \Eprint
  {https://arxiv.org/abs/https://doi.org/10.1021/acs.nanolett.9b02298} {https://doi.org/10.1021/acs.nanolett.9b02298} \BibitemShut {NoStop}%
\bibitem [{\citenamefont {Thiel}\ \emph {et~al.}(2016)\citenamefont {Thiel}, \citenamefont {Rohner}, \citenamefont {Ganzhorn}, \citenamefont {Appel}, \citenamefont {Neu}, \citenamefont {M{\"u}ller}, \citenamefont {Kleiner}, \citenamefont {Koelle},\ and\ \citenamefont {Maletinsky}}]{Thiel2016Quantative}%
  \BibitemOpen
  \bibfield  {author} {\bibinfo {author} {\bibfnamefont {L.}~\bibnamefont {Thiel}}, \bibinfo {author} {\bibfnamefont {D.}~\bibnamefont {Rohner}}, \bibinfo {author} {\bibfnamefont {M.}~\bibnamefont {Ganzhorn}}, \bibinfo {author} {\bibfnamefont {P.}~\bibnamefont {Appel}}, \bibinfo {author} {\bibfnamefont {E.}~\bibnamefont {Neu}}, \bibinfo {author} {\bibfnamefont {B.}~\bibnamefont {M{\"u}ller}}, \bibinfo {author} {\bibfnamefont {R.}~\bibnamefont {Kleiner}}, \bibinfo {author} {\bibfnamefont {D.}~\bibnamefont {Koelle}},\ and\ \bibinfo {author} {\bibfnamefont {P.}~\bibnamefont {Maletinsky}},\ }\bibfield  {title} {\bibinfo {title} {Quantitative nanoscale vortex imaging using a cryogenic quantum magnetometer},\ }\href {https://doi.org/10.1038/nnano.2016.63} {\bibfield  {journal} {\bibinfo  {journal} {Nature Nanotechnology}\ }\textbf {\bibinfo {volume} {11}},\ \bibinfo {pages} {677} (\bibinfo {year} {2016})}\BibitemShut {NoStop}%
\bibitem [{\citenamefont {Pelliccione}\ \emph {et~al.}(2016)\citenamefont {Pelliccione}, \citenamefont {Jenkins}, \citenamefont {Ovartchaiyapong}, \citenamefont {Reetz}, \citenamefont {Emmanouilidou}, \citenamefont {Ni},\ and\ \citenamefont {Bleszynski~Jayich}}]{Pelliccione2016Scanned}%
  \BibitemOpen
  \bibfield  {author} {\bibinfo {author} {\bibfnamefont {M.}~\bibnamefont {Pelliccione}}, \bibinfo {author} {\bibfnamefont {A.}~\bibnamefont {Jenkins}}, \bibinfo {author} {\bibfnamefont {P.}~\bibnamefont {Ovartchaiyapong}}, \bibinfo {author} {\bibfnamefont {C.}~\bibnamefont {Reetz}}, \bibinfo {author} {\bibfnamefont {E.}~\bibnamefont {Emmanouilidou}}, \bibinfo {author} {\bibfnamefont {N.}~\bibnamefont {Ni}},\ and\ \bibinfo {author} {\bibfnamefont {A.~C.}\ \bibnamefont {Bleszynski~Jayich}},\ }\bibfield  {title} {\bibinfo {title} {Scanned probe imaging of nanoscale magnetism at cryogenic temperatures with a single-spin quantum sensor},\ }\href {https://doi.org/10.1038/nnano.2016.68} {\bibfield  {journal} {\bibinfo  {journal} {Nature Nanotechnology}\ }\textbf {\bibinfo {volume} {11}},\ \bibinfo {pages} {700} (\bibinfo {year} {2016})}\BibitemShut {NoStop}%
\bibitem [{\citenamefont {Schlussel}\ \emph {et~al.}(2018)\citenamefont {Schlussel}, \citenamefont {Lenz}, \citenamefont {Rohner}, \citenamefont {Bar-Haim}, \citenamefont {Bougas}, \citenamefont {Groswasser}, \citenamefont {Kieschnick}, \citenamefont {Rozenberg}, \citenamefont {Thiel}, \citenamefont {Waxman}, \citenamefont {Meijer}, \citenamefont {Maletinsky}, \citenamefont {Budker},\ and\ \citenamefont {Folman}}]{Schlussel2018Wide}%
  \BibitemOpen
  \bibfield  {author} {\bibinfo {author} {\bibfnamefont {Y.}~\bibnamefont {Schlussel}}, \bibinfo {author} {\bibfnamefont {T.}~\bibnamefont {Lenz}}, \bibinfo {author} {\bibfnamefont {D.}~\bibnamefont {Rohner}}, \bibinfo {author} {\bibfnamefont {Y.}~\bibnamefont {Bar-Haim}}, \bibinfo {author} {\bibfnamefont {L.}~\bibnamefont {Bougas}}, \bibinfo {author} {\bibfnamefont {D.}~\bibnamefont {Groswasser}}, \bibinfo {author} {\bibfnamefont {M.}~\bibnamefont {Kieschnick}}, \bibinfo {author} {\bibfnamefont {E.}~\bibnamefont {Rozenberg}}, \bibinfo {author} {\bibfnamefont {L.}~\bibnamefont {Thiel}}, \bibinfo {author} {\bibfnamefont {A.}~\bibnamefont {Waxman}}, \bibinfo {author} {\bibfnamefont {J.}~\bibnamefont {Meijer}}, \bibinfo {author} {\bibfnamefont {P.}~\bibnamefont {Maletinsky}}, \bibinfo {author} {\bibfnamefont {D.}~\bibnamefont {Budker}},\ and\ \bibinfo {author} {\bibfnamefont {R.}~\bibnamefont {Folman}},\ }\bibfield  {title} {\bibinfo {title} {Wide-field imaging of superconductor vortices with electron spins in
  diamond},\ }\href {https://doi.org/10.1103/PhysRevApplied.10.034032} {\bibfield  {journal} {\bibinfo  {journal} {Phys. Rev. Applied}\ }\textbf {\bibinfo {volume} {10}},\ \bibinfo {pages} {034032} (\bibinfo {year} {2018})}\BibitemShut {NoStop}%
\bibitem [{\citenamefont {Ho}\ \emph {et~al.}(2021{\natexlab{a}})\citenamefont {Ho}, \citenamefont {Shen}, \citenamefont {Pang}, \citenamefont {Leung}, \citenamefont {Zhao},\ and\ \citenamefont {Yang}}]{Ho2021Diamond}%
  \BibitemOpen
  \bibfield  {author} {\bibinfo {author} {\bibfnamefont {K.~O.}\ \bibnamefont {Ho}}, \bibinfo {author} {\bibfnamefont {Y.}~\bibnamefont {Shen}}, \bibinfo {author} {\bibfnamefont {Y.~Y.}\ \bibnamefont {Pang}}, \bibinfo {author} {\bibfnamefont {W.~K.}\ \bibnamefont {Leung}}, \bibinfo {author} {\bibfnamefont {N.}~\bibnamefont {Zhao}},\ and\ \bibinfo {author} {\bibfnamefont {S.}~\bibnamefont {Yang}},\ }\bibfield  {title} {\bibinfo {title} {Diamond quantum sensors: from physics to applications on condensed matter research},\ }\href {https://doi.org/10.1080/26941112.2021.1964926} {\bibfield  {journal} {\bibinfo  {journal} {Functional Diamond}\ }\textbf {\bibinfo {volume} {1}},\ \bibinfo {pages} {160} (\bibinfo {year} {2021}{\natexlab{a}})},\ \Eprint {https://arxiv.org/abs/https://doi.org/10.1080/26941112.2021.1964926} {https://doi.org/10.1080/26941112.2021.1964926} \BibitemShut {NoStop}%
\bibitem [{\citenamefont {McLaughlin}\ \emph {et~al.}(2021)\citenamefont {McLaughlin}, \citenamefont {Wang}, \citenamefont {Huang}, \citenamefont {Lee-Wong}, \citenamefont {Hu}, \citenamefont {Lu}, \citenamefont {Yan}, \citenamefont {Gu}, \citenamefont {Wu}, \citenamefont {You},\ and\ \citenamefont {Du}}]{McLaughlin2021Strong}%
  \BibitemOpen
  \bibfield  {author} {\bibinfo {author} {\bibfnamefont {N.~J.}\ \bibnamefont {McLaughlin}}, \bibinfo {author} {\bibfnamefont {H.}~\bibnamefont {Wang}}, \bibinfo {author} {\bibfnamefont {M.}~\bibnamefont {Huang}}, \bibinfo {author} {\bibfnamefont {E.}~\bibnamefont {Lee-Wong}}, \bibinfo {author} {\bibfnamefont {L.}~\bibnamefont {Hu}}, \bibinfo {author} {\bibfnamefont {H.}~\bibnamefont {Lu}}, \bibinfo {author} {\bibfnamefont {G.~Q.}\ \bibnamefont {Yan}}, \bibinfo {author} {\bibfnamefont {G.}~\bibnamefont {Gu}}, \bibinfo {author} {\bibfnamefont {C.}~\bibnamefont {Wu}}, \bibinfo {author} {\bibfnamefont {Y.-Z.}\ \bibnamefont {You}},\ and\ \bibinfo {author} {\bibfnamefont {C.~R.}\ \bibnamefont {Du}},\ }\bibfield  {title} {\bibinfo {title} {Strong correlation between superconductivity and ferromagnetism in an fe-chalcogenide superconductor},\ }\href {https://doi.org/10.1021/acs.nanolett.1c02424} {\bibfield  {journal} {\bibinfo  {journal} {Nano Letters}\ }\textbf {\bibinfo {volume} {21}},\ \bibinfo {pages} {7277}
  (\bibinfo {year} {2021})},\ \bibinfo {note} {pMID: 34415171},\ \Eprint {https://arxiv.org/abs/https://doi.org/10.1021/acs.nanolett.1c02424} {https://doi.org/10.1021/acs.nanolett.1c02424} \BibitemShut {NoStop}%
\bibitem [{\citenamefont {Monge}\ \emph {et~al.}(2023)\citenamefont {Monge}, \citenamefont {Delord}, \citenamefont {Proscia}, \citenamefont {Shotan}, \citenamefont {Jayakumar}, \citenamefont {Henshaw}, \citenamefont {Zangara}, \citenamefont {Lozovoi}, \citenamefont {Pagliero}, \citenamefont {Esquinazi}, \citenamefont {An}, \citenamefont {Sodemann}, \citenamefont {Menon},\ and\ \citenamefont {Meriles}}]{Monge2023Spin}%
  \BibitemOpen
  \bibfield  {author} {\bibinfo {author} {\bibfnamefont {R.}~\bibnamefont {Monge}}, \bibinfo {author} {\bibfnamefont {T.}~\bibnamefont {Delord}}, \bibinfo {author} {\bibfnamefont {N.~V.}\ \bibnamefont {Proscia}}, \bibinfo {author} {\bibfnamefont {Z.}~\bibnamefont {Shotan}}, \bibinfo {author} {\bibfnamefont {H.}~\bibnamefont {Jayakumar}}, \bibinfo {author} {\bibfnamefont {J.}~\bibnamefont {Henshaw}}, \bibinfo {author} {\bibfnamefont {P.~R.}\ \bibnamefont {Zangara}}, \bibinfo {author} {\bibfnamefont {A.}~\bibnamefont {Lozovoi}}, \bibinfo {author} {\bibfnamefont {D.}~\bibnamefont {Pagliero}}, \bibinfo {author} {\bibfnamefont {P.~D.}\ \bibnamefont {Esquinazi}}, \bibinfo {author} {\bibfnamefont {T.}~\bibnamefont {An}}, \bibinfo {author} {\bibfnamefont {I.}~\bibnamefont {Sodemann}}, \bibinfo {author} {\bibfnamefont {V.~M.}\ \bibnamefont {Menon}},\ and\ \bibinfo {author} {\bibfnamefont {C.~A.}\ \bibnamefont {Meriles}},\ }\bibfield  {title} {\bibinfo {title} {Spin dynamics of a solid-state qubit in proximity to a
  superconductor},\ }\href {https://doi.org/10.1021/acs.nanolett.2c03250} {\bibfield  {journal} {\bibinfo  {journal} {Nano Letters}\ }\textbf {\bibinfo {volume} {23}},\ \bibinfo {pages} {422} (\bibinfo {year} {2023})},\ \bibinfo {note} {pMID: 36602464},\ \Eprint {https://arxiv.org/abs/https://doi.org/10.1021/acs.nanolett.2c03250} {https://doi.org/10.1021/acs.nanolett.2c03250} \BibitemShut {NoStop}%
\bibitem [{\citenamefont {Casola}\ \emph {et~al.}(2018)\citenamefont {Casola}, \citenamefont {van~der Sar},\ and\ \citenamefont {Yacoby}}]{Casola2018Probing}%
  \BibitemOpen
  \bibfield  {author} {\bibinfo {author} {\bibfnamefont {F.}~\bibnamefont {Casola}}, \bibinfo {author} {\bibfnamefont {T.}~\bibnamefont {van~der Sar}},\ and\ \bibinfo {author} {\bibfnamefont {A.}~\bibnamefont {Yacoby}},\ }\bibfield  {title} {\bibinfo {title} {Probing condensed matter physics with magnetometry based on nitrogen-vacancy centres in diamond},\ }\href {https://doi.org/10.1038/natrevmats.2017.88} {\bibfield  {journal} {\bibinfo  {journal} {Nature Reviews Materials}\ }\textbf {\bibinfo {volume} {3}},\ \bibinfo {pages} {17088} (\bibinfo {year} {2018})}\BibitemShut {NoStop}%
\bibitem [{\citenamefont {Lillie}\ \emph {et~al.}(2020)\citenamefont {Lillie}, \citenamefont {Broadway}, \citenamefont {Dontschuk}, \citenamefont {Scholten}, \citenamefont {Johnson}, \citenamefont {Wolf}, \citenamefont {Rachel}, \citenamefont {Hollenberg},\ and\ \citenamefont {Tetienne}}]{Lillie2020Laser}%
  \BibitemOpen
  \bibfield  {author} {\bibinfo {author} {\bibfnamefont {S.~E.}\ \bibnamefont {Lillie}}, \bibinfo {author} {\bibfnamefont {D.~A.}\ \bibnamefont {Broadway}}, \bibinfo {author} {\bibfnamefont {N.}~\bibnamefont {Dontschuk}}, \bibinfo {author} {\bibfnamefont {S.~C.}\ \bibnamefont {Scholten}}, \bibinfo {author} {\bibfnamefont {B.~C.}\ \bibnamefont {Johnson}}, \bibinfo {author} {\bibfnamefont {S.}~\bibnamefont {Wolf}}, \bibinfo {author} {\bibfnamefont {S.}~\bibnamefont {Rachel}}, \bibinfo {author} {\bibfnamefont {L.~C.~L.}\ \bibnamefont {Hollenberg}},\ and\ \bibinfo {author} {\bibfnamefont {J.-P.}\ \bibnamefont {Tetienne}},\ }\bibfield  {title} {\bibinfo {title} {Laser modulation of superconductivity in a cryogenic wide-field nitrogen-vacancy microscope},\ }\href {https://doi.org/10.1021/acs.nanolett.9b05071} {\bibfield  {journal} {\bibinfo  {journal} {Nano Letters}\ }\textbf {\bibinfo {volume} {20}},\ \bibinfo {pages} {1855} (\bibinfo {year} {2020})},\ \bibinfo {note} {pMID: 32017577},\ \Eprint
  {https://arxiv.org/abs/https://doi.org/10.1021/acs.nanolett.9b05071} {https://doi.org/10.1021/acs.nanolett.9b05071} \BibitemShut {NoStop}%
\bibitem [{\citenamefont {Rohner}\ \emph {et~al.}(2018)\citenamefont {Rohner}, \citenamefont {Thiel}, \citenamefont {Müller}, \citenamefont {Kasperczyk}, \citenamefont {Kleiner}, \citenamefont {Koelle},\ and\ \citenamefont {Maletinsky}}]{Rohner2018Real}%
  \BibitemOpen
  \bibfield  {author} {\bibinfo {author} {\bibfnamefont {D.}~\bibnamefont {Rohner}}, \bibinfo {author} {\bibfnamefont {L.}~\bibnamefont {Thiel}}, \bibinfo {author} {\bibfnamefont {B.}~\bibnamefont {Müller}}, \bibinfo {author} {\bibfnamefont {M.}~\bibnamefont {Kasperczyk}}, \bibinfo {author} {\bibfnamefont {R.}~\bibnamefont {Kleiner}}, \bibinfo {author} {\bibfnamefont {D.}~\bibnamefont {Koelle}},\ and\ \bibinfo {author} {\bibfnamefont {P.}~\bibnamefont {Maletinsky}},\ }\bibfield  {title} {\bibinfo {title} {Real-space probing of the local magnetic response of thin-film superconductors using single spin magnetometry},\ }\bibfield  {journal} {\bibinfo  {journal} {Sensors}\ }\textbf {\bibinfo {volume} {18}},\ \href {https://doi.org/10.3390/s18113790} {10.3390/s18113790} (\bibinfo {year} {2018})\BibitemShut {NoStop}%
\bibitem [{\citenamefont {Scheidegger}\ \emph {et~al.}(2022)\citenamefont {Scheidegger}, \citenamefont {Diesch}, \citenamefont {Palm},\ and\ \citenamefont {Degen}}]{Scheidegger2022Scanning}%
  \BibitemOpen
  \bibfield  {author} {\bibinfo {author} {\bibfnamefont {P.~J.}\ \bibnamefont {Scheidegger}}, \bibinfo {author} {\bibfnamefont {S.}~\bibnamefont {Diesch}}, \bibinfo {author} {\bibfnamefont {M.~L.}\ \bibnamefont {Palm}},\ and\ \bibinfo {author} {\bibfnamefont {C.~L.}\ \bibnamefont {Degen}},\ }\bibfield  {title} {\bibinfo {title} {Scanning nitrogen-vacancy magnetometry down to 350 mk},\ }\href {https://doi.org/10.1063/5.0093548} {\bibfield  {journal} {\bibinfo  {journal} {Applied Physics Letters}\ }\textbf {\bibinfo {volume} {120}},\ \bibinfo {pages} {224001} (\bibinfo {year} {2022})},\ \Eprint {https://arxiv.org/abs/https://doi.org/10.1063/5.0093548} {https://doi.org/10.1063/5.0093548} \BibitemShut {NoStop}%
\bibitem [{\citenamefont {Paone}\ \emph {et~al.}(2021)\citenamefont {Paone}, \citenamefont {Pinto}, \citenamefont {Kim}, \citenamefont {Feng}, \citenamefont {Kim}, \citenamefont {Stöhr}, \citenamefont {Singha}, \citenamefont {Kaiser}, \citenamefont {Logvenov}, \citenamefont {Keimer}, \citenamefont {Wrachtrup},\ and\ \citenamefont {Kern}}]{Paone2021All}%
  \BibitemOpen
  \bibfield  {author} {\bibinfo {author} {\bibfnamefont {D.}~\bibnamefont {Paone}}, \bibinfo {author} {\bibfnamefont {D.}~\bibnamefont {Pinto}}, \bibinfo {author} {\bibfnamefont {G.}~\bibnamefont {Kim}}, \bibinfo {author} {\bibfnamefont {L.}~\bibnamefont {Feng}}, \bibinfo {author} {\bibfnamefont {M.-J.}\ \bibnamefont {Kim}}, \bibinfo {author} {\bibfnamefont {R.}~\bibnamefont {Stöhr}}, \bibinfo {author} {\bibfnamefont {A.}~\bibnamefont {Singha}}, \bibinfo {author} {\bibfnamefont {S.}~\bibnamefont {Kaiser}}, \bibinfo {author} {\bibfnamefont {G.}~\bibnamefont {Logvenov}}, \bibinfo {author} {\bibfnamefont {B.}~\bibnamefont {Keimer}}, \bibinfo {author} {\bibfnamefont {J.}~\bibnamefont {Wrachtrup}},\ and\ \bibinfo {author} {\bibfnamefont {K.}~\bibnamefont {Kern}},\ }\bibfield  {title} {\bibinfo {title} {All-optical and microwave-free detection of meissner screening using nitrogen-vacancy centers in diamond},\ }\href {https://doi.org/10.1063/5.0037414} {\bibfield  {journal} {\bibinfo  {journal} {Journal of Applied
  Physics}\ }\textbf {\bibinfo {volume} {129}},\ \bibinfo {pages} {024306} (\bibinfo {year} {2021})},\ \Eprint {https://arxiv.org/abs/https://doi.org/10.1063/5.0037414} {https://doi.org/10.1063/5.0037414} \BibitemShut {NoStop}%
\bibitem [{\citenamefont {Joshi}\ \emph {et~al.}(2020)\citenamefont {Joshi}, \citenamefont {Nusran}, \citenamefont {Tanatar}, \citenamefont {Cho}, \citenamefont {Bud’ko}, \citenamefont {Canfield}, \citenamefont {Fernandes}, \citenamefont {Levchenko},\ and\ \citenamefont {Prozorov}}]{Joshi2020Quantum}%
  \BibitemOpen
  \bibfield  {author} {\bibinfo {author} {\bibfnamefont {K.~R.}\ \bibnamefont {Joshi}}, \bibinfo {author} {\bibfnamefont {N.~M.}\ \bibnamefont {Nusran}}, \bibinfo {author} {\bibfnamefont {M.~A.}\ \bibnamefont {Tanatar}}, \bibinfo {author} {\bibfnamefont {K.}~\bibnamefont {Cho}}, \bibinfo {author} {\bibfnamefont {S.~L.}\ \bibnamefont {Bud’ko}}, \bibinfo {author} {\bibfnamefont {P.~C.}\ \bibnamefont {Canfield}}, \bibinfo {author} {\bibfnamefont {R.~M.}\ \bibnamefont {Fernandes}}, \bibinfo {author} {\bibfnamefont {A.}~\bibnamefont {Levchenko}},\ and\ \bibinfo {author} {\bibfnamefont {R.}~\bibnamefont {Prozorov}},\ }\bibfield  {title} {\bibinfo {title} {Quantum phase transition inside the superconducting dome of ba(fe1-xcox)2as2 from diamond-based optical magnetometry},\ }\href {https://doi.org/10.1088/1367-2630/ab85a9} {\bibfield  {journal} {\bibinfo  {journal} {New Journal of Physics}\ }\textbf {\bibinfo {volume} {22}},\ \bibinfo {pages} {053037} (\bibinfo {year} {2020})}\BibitemShut {NoStop}%
\bibitem [{\citenamefont {Yip}\ \emph {et~al.}(2019)\citenamefont {Yip}, \citenamefont {Ho}, \citenamefont {Yu}, \citenamefont {Chen}, \citenamefont {Zhang}, \citenamefont {Kasahara}, \citenamefont {Mizukami}, \citenamefont {Shibauchi}, \citenamefont {Matsuda}, \citenamefont {Goh},\ and\ \citenamefont {Yang}}]{Yip2019Measuring}%
  \BibitemOpen
  \bibfield  {author} {\bibinfo {author} {\bibfnamefont {K.~Y.}\ \bibnamefont {Yip}}, \bibinfo {author} {\bibfnamefont {K.~O.}\ \bibnamefont {Ho}}, \bibinfo {author} {\bibfnamefont {K.~Y.}\ \bibnamefont {Yu}}, \bibinfo {author} {\bibfnamefont {Y.}~\bibnamefont {Chen}}, \bibinfo {author} {\bibfnamefont {W.}~\bibnamefont {Zhang}}, \bibinfo {author} {\bibfnamefont {S.}~\bibnamefont {Kasahara}}, \bibinfo {author} {\bibfnamefont {Y.}~\bibnamefont {Mizukami}}, \bibinfo {author} {\bibfnamefont {T.}~\bibnamefont {Shibauchi}}, \bibinfo {author} {\bibfnamefont {Y.}~\bibnamefont {Matsuda}}, \bibinfo {author} {\bibfnamefont {S.~K.}\ \bibnamefont {Goh}},\ and\ \bibinfo {author} {\bibfnamefont {S.}~\bibnamefont {Yang}},\ }\bibfield  {title} {\bibinfo {title} {Measuring magnetic field texture in correlated electron systems under extreme conditions},\ }\href {https://doi.org/10.1126/science.aaw4278} {\bibfield  {journal} {\bibinfo  {journal} {Science}\ }\textbf {\bibinfo {volume} {366}},\ \bibinfo {pages} {1355} (\bibinfo
  {year} {2019})},\ \Eprint {https://arxiv.org/abs/https://science.sciencemag.org/content/366/6471/1355.full.pdf} {https://science.sciencemag.org/content/366/6471/1355.full.pdf} \BibitemShut {NoStop}%
\bibitem [{\citenamefont {Lesik}\ \emph {et~al.}(2019)\citenamefont {Lesik}, \citenamefont {Plisson}, \citenamefont {Toraille}, \citenamefont {Renaud}, \citenamefont {Occelli}, \citenamefont {Schmidt}, \citenamefont {Salord}, \citenamefont {Delobbe}, \citenamefont {Debuisschert}, \citenamefont {Rondin}, \citenamefont {Loubeyre},\ and\ \citenamefont {Roch}}]{Lesik2019Magnetic}%
  \BibitemOpen
  \bibfield  {author} {\bibinfo {author} {\bibfnamefont {M.}~\bibnamefont {Lesik}}, \bibinfo {author} {\bibfnamefont {T.}~\bibnamefont {Plisson}}, \bibinfo {author} {\bibfnamefont {L.}~\bibnamefont {Toraille}}, \bibinfo {author} {\bibfnamefont {J.}~\bibnamefont {Renaud}}, \bibinfo {author} {\bibfnamefont {F.}~\bibnamefont {Occelli}}, \bibinfo {author} {\bibfnamefont {M.}~\bibnamefont {Schmidt}}, \bibinfo {author} {\bibfnamefont {O.}~\bibnamefont {Salord}}, \bibinfo {author} {\bibfnamefont {A.}~\bibnamefont {Delobbe}}, \bibinfo {author} {\bibfnamefont {T.}~\bibnamefont {Debuisschert}}, \bibinfo {author} {\bibfnamefont {L.}~\bibnamefont {Rondin}}, \bibinfo {author} {\bibfnamefont {P.}~\bibnamefont {Loubeyre}},\ and\ \bibinfo {author} {\bibfnamefont {J.-F.}\ \bibnamefont {Roch}},\ }\bibfield  {title} {\bibinfo {title} {Magnetic measurements on micrometer-sized samples under high pressure using designed nv centers},\ }\href {https://doi.org/10.1126/science.aaw4329} {\bibfield  {journal} {\bibinfo  {journal}
  {Science}\ }\textbf {\bibinfo {volume} {366}},\ \bibinfo {pages} {1359} (\bibinfo {year} {2019})},\ \Eprint {https://arxiv.org/abs/https://science.sciencemag.org/content/366/6471/1359.full.pdf} {https://science.sciencemag.org/content/366/6471/1359.full.pdf} \BibitemShut {NoStop}%
\bibitem [{\citenamefont {Hsieh}\ \emph {et~al.}(2019)\citenamefont {Hsieh}, \citenamefont {Bhattacharyya}, \citenamefont {Zu}, \citenamefont {Mittiga}, \citenamefont {Smart}, \citenamefont {Machado}, \citenamefont {Kobrin}, \citenamefont {H{\"o}hn}, \citenamefont {Rui}, \citenamefont {Kamrani}, \citenamefont {Chatterjee}, \citenamefont {Choi}, \citenamefont {Zaletel}, \citenamefont {Struzhkin}, \citenamefont {Moore}, \citenamefont {Levitas}, \citenamefont {Jeanloz},\ and\ \citenamefont {Yao}}]{Hsieh2019Imaging}%
  \BibitemOpen
  \bibfield  {author} {\bibinfo {author} {\bibfnamefont {S.}~\bibnamefont {Hsieh}}, \bibinfo {author} {\bibfnamefont {P.}~\bibnamefont {Bhattacharyya}}, \bibinfo {author} {\bibfnamefont {C.}~\bibnamefont {Zu}}, \bibinfo {author} {\bibfnamefont {T.}~\bibnamefont {Mittiga}}, \bibinfo {author} {\bibfnamefont {T.~J.}\ \bibnamefont {Smart}}, \bibinfo {author} {\bibfnamefont {F.}~\bibnamefont {Machado}}, \bibinfo {author} {\bibfnamefont {B.}~\bibnamefont {Kobrin}}, \bibinfo {author} {\bibfnamefont {T.~O.}\ \bibnamefont {H{\"o}hn}}, \bibinfo {author} {\bibfnamefont {N.~Z.}\ \bibnamefont {Rui}}, \bibinfo {author} {\bibfnamefont {M.}~\bibnamefont {Kamrani}}, \bibinfo {author} {\bibfnamefont {S.}~\bibnamefont {Chatterjee}}, \bibinfo {author} {\bibfnamefont {S.}~\bibnamefont {Choi}}, \bibinfo {author} {\bibfnamefont {M.}~\bibnamefont {Zaletel}}, \bibinfo {author} {\bibfnamefont {V.~V.}\ \bibnamefont {Struzhkin}}, \bibinfo {author} {\bibfnamefont {J.~E.}\ \bibnamefont {Moore}}, \bibinfo {author} {\bibfnamefont {V.~I.}\
  \bibnamefont {Levitas}}, \bibinfo {author} {\bibfnamefont {R.}~\bibnamefont {Jeanloz}},\ and\ \bibinfo {author} {\bibfnamefont {N.~Y.}\ \bibnamefont {Yao}},\ }\bibfield  {title} {\bibinfo {title} {Imaging stress and magnetism at high pressures using a nanoscale quantum sensor},\ }\href {https://doi.org/10.1126/science.aaw4352} {\bibfield  {journal} {\bibinfo  {journal} {Science}\ }\textbf {\bibinfo {volume} {366}},\ \bibinfo {pages} {1349} (\bibinfo {year} {2019})},\ \Eprint {https://arxiv.org/abs/https://science.sciencemag.org/content/366/6471/1349.full.pdf} {https://science.sciencemag.org/content/366/6471/1349.full.pdf} \BibitemShut {NoStop}%
\bibitem [{\citenamefont {Ho}\ \emph {et~al.}(2020)\citenamefont {Ho}, \citenamefont {Leung}, \citenamefont {Jiang}, \citenamefont {Ao}, \citenamefont {Zhang}, \citenamefont {Yip}, \citenamefont {Pang}, \citenamefont {Wong}, \citenamefont {Goh},\ and\ \citenamefont {Yang}}]{Ho2020Probing}%
  \BibitemOpen
  \bibfield  {author} {\bibinfo {author} {\bibfnamefont {K.~O.}\ \bibnamefont {Ho}}, \bibinfo {author} {\bibfnamefont {M.~Y.}\ \bibnamefont {Leung}}, \bibinfo {author} {\bibfnamefont {Y.}~\bibnamefont {Jiang}}, \bibinfo {author} {\bibfnamefont {K.~P.}\ \bibnamefont {Ao}}, \bibinfo {author} {\bibfnamefont {W.}~\bibnamefont {Zhang}}, \bibinfo {author} {\bibfnamefont {K.~Y.}\ \bibnamefont {Yip}}, \bibinfo {author} {\bibfnamefont {Y.~Y.}\ \bibnamefont {Pang}}, \bibinfo {author} {\bibfnamefont {K.~C.}\ \bibnamefont {Wong}}, \bibinfo {author} {\bibfnamefont {S.~K.}\ \bibnamefont {Goh}},\ and\ \bibinfo {author} {\bibfnamefont {S.}~\bibnamefont {Yang}},\ }\bibfield  {title} {\bibinfo {title} {Probing local pressure environment in anvil cells with nitrogen-vacancy (n-${V}^{\ensuremath{-}}$) centers in diamond},\ }\href {https://doi.org/10.1103/PhysRevApplied.13.024041} {\bibfield  {journal} {\bibinfo  {journal} {Phys. Rev. Applied}\ }\textbf {\bibinfo {volume} {13}},\ \bibinfo {pages} {024041} (\bibinfo {year}
  {2020})}\BibitemShut {NoStop}%
\bibitem [{\citenamefont {Ho}\ \emph {et~al.}(2021{\natexlab{b}})\citenamefont {Ho}, \citenamefont {Wong}, \citenamefont {Leung}, \citenamefont {Pang}, \citenamefont {Leung}, \citenamefont {Yip}, \citenamefont {Zhang}, \citenamefont {Xie}, \citenamefont {Goh},\ and\ \citenamefont {Yang}}]{Ho2021Recent}%
  \BibitemOpen
  \bibfield  {author} {\bibinfo {author} {\bibfnamefont {K.~O.}\ \bibnamefont {Ho}}, \bibinfo {author} {\bibfnamefont {K.~C.}\ \bibnamefont {Wong}}, \bibinfo {author} {\bibfnamefont {M.~Y.}\ \bibnamefont {Leung}}, \bibinfo {author} {\bibfnamefont {Y.~Y.}\ \bibnamefont {Pang}}, \bibinfo {author} {\bibfnamefont {W.~K.}\ \bibnamefont {Leung}}, \bibinfo {author} {\bibfnamefont {K.~Y.}\ \bibnamefont {Yip}}, \bibinfo {author} {\bibfnamefont {W.}~\bibnamefont {Zhang}}, \bibinfo {author} {\bibfnamefont {J.}~\bibnamefont {Xie}}, \bibinfo {author} {\bibfnamefont {S.~K.}\ \bibnamefont {Goh}},\ and\ \bibinfo {author} {\bibfnamefont {S.}~\bibnamefont {Yang}},\ }\bibfield  {title} {\bibinfo {title} {Recent developments of quantum sensing under pressurized environment using the nitrogen vacancy (nv) center in diamond},\ }\href {https://doi.org/10.1063/5.0052233} {\bibfield  {journal} {\bibinfo  {journal} {Journal of Applied Physics}\ }\textbf {\bibinfo {volume} {129}},\ \bibinfo {pages} {241101} (\bibinfo {year}
  {2021}{\natexlab{b}})},\ \Eprint {https://arxiv.org/abs/https://doi.org/10.1063/5.0052233} {https://doi.org/10.1063/5.0052233} \BibitemShut {NoStop}%
\bibitem [{\citenamefont {Toraille}\ \emph {et~al.}(2020)\citenamefont {Toraille}, \citenamefont {Hilberer}, \citenamefont {Plisson}, \citenamefont {Lesik}, \citenamefont {Chipaux}, \citenamefont {Vindolet}, \citenamefont {Pépin}, \citenamefont {Occelli}, \citenamefont {Schmidt}, \citenamefont {Debuisschert}, \citenamefont {Guignot}, \citenamefont {Itié}, \citenamefont {Loubeyre},\ and\ \citenamefont {Roch}}]{Toraille2020Combined}%
  \BibitemOpen
  \bibfield  {author} {\bibinfo {author} {\bibfnamefont {L.}~\bibnamefont {Toraille}}, \bibinfo {author} {\bibfnamefont {A.}~\bibnamefont {Hilberer}}, \bibinfo {author} {\bibfnamefont {T.}~\bibnamefont {Plisson}}, \bibinfo {author} {\bibfnamefont {M.}~\bibnamefont {Lesik}}, \bibinfo {author} {\bibfnamefont {M.}~\bibnamefont {Chipaux}}, \bibinfo {author} {\bibfnamefont {B.}~\bibnamefont {Vindolet}}, \bibinfo {author} {\bibfnamefont {C.}~\bibnamefont {Pépin}}, \bibinfo {author} {\bibfnamefont {F.}~\bibnamefont {Occelli}}, \bibinfo {author} {\bibfnamefont {M.}~\bibnamefont {Schmidt}}, \bibinfo {author} {\bibfnamefont {T.}~\bibnamefont {Debuisschert}}, \bibinfo {author} {\bibfnamefont {N.}~\bibnamefont {Guignot}}, \bibinfo {author} {\bibfnamefont {J.-P.}\ \bibnamefont {Itié}}, \bibinfo {author} {\bibfnamefont {P.}~\bibnamefont {Loubeyre}},\ and\ \bibinfo {author} {\bibfnamefont {J.-F.}\ \bibnamefont {Roch}},\ }\bibfield  {title} {\bibinfo {title} {Combined synchrotron x-ray diffraction and nv diamond magnetic
  microscopy measurements at high pressure},\ }\href {https://doi.org/10.1088/1367-2630/abc28f} {\bibfield  {journal} {\bibinfo  {journal} {New Journal of Physics}\ }\textbf {\bibinfo {volume} {22}},\ \bibinfo {pages} {103063} (\bibinfo {year} {2020})}\BibitemShut {NoStop}%
\bibitem [{\citenamefont {Ho}\ \emph {et~al.}(2023)\citenamefont {Ho}, \citenamefont {Leung}, \citenamefont {Wang}, \citenamefont {Xie}, \citenamefont {Yip}, \citenamefont {Wu}, \citenamefont {Goh}, \citenamefont {Denisenko}, \citenamefont {Wrachtrup},\ and\ \citenamefont {Yang}}]{Ho2023Spectroscopic}%
  \BibitemOpen
  \bibfield  {author} {\bibinfo {author} {\bibfnamefont {K.~O.}\ \bibnamefont {Ho}}, \bibinfo {author} {\bibfnamefont {M.~Y.}\ \bibnamefont {Leung}}, \bibinfo {author} {\bibfnamefont {W.}~\bibnamefont {Wang}}, \bibinfo {author} {\bibfnamefont {J.}~\bibnamefont {Xie}}, \bibinfo {author} {\bibfnamefont {K.~Y.}\ \bibnamefont {Yip}}, \bibinfo {author} {\bibfnamefont {J.}~\bibnamefont {Wu}}, \bibinfo {author} {\bibfnamefont {S.~K.}\ \bibnamefont {Goh}}, \bibinfo {author} {\bibfnamefont {A.}~\bibnamefont {Denisenko}}, \bibinfo {author} {\bibfnamefont {J.}~\bibnamefont {Wrachtrup}},\ and\ \bibinfo {author} {\bibfnamefont {S.}~\bibnamefont {Yang}},\ }\bibfield  {title} {\bibinfo {title} {Spectroscopic study of n-$v$ sensors in diamond-based high-pressure devices},\ }\href {https://doi.org/10.1103/PhysRevApplied.19.044091} {\bibfield  {journal} {\bibinfo  {journal} {Phys. Rev. Appl.}\ }\textbf {\bibinfo {volume} {19}},\ \bibinfo {pages} {044091} (\bibinfo {year} {2023})}\BibitemShut {NoStop}%
\bibitem [{\citenamefont {Hilberer}\ \emph {et~al.}(2023)\citenamefont {Hilberer}, \citenamefont {Toraille}, \citenamefont {Dailledouze}, \citenamefont {Adam}, \citenamefont {Hanlon}, \citenamefont {Weck}, \citenamefont {Schmidt}, \citenamefont {Loubeyre},\ and\ \citenamefont {Roch}}]{Hilberer2023Enabling}%
  \BibitemOpen
  \bibfield  {author} {\bibinfo {author} {\bibfnamefont {A.}~\bibnamefont {Hilberer}}, \bibinfo {author} {\bibfnamefont {L.}~\bibnamefont {Toraille}}, \bibinfo {author} {\bibfnamefont {C.}~\bibnamefont {Dailledouze}}, \bibinfo {author} {\bibfnamefont {M.-P.}\ \bibnamefont {Adam}}, \bibinfo {author} {\bibfnamefont {L.}~\bibnamefont {Hanlon}}, \bibinfo {author} {\bibfnamefont {G.}~\bibnamefont {Weck}}, \bibinfo {author} {\bibfnamefont {M.}~\bibnamefont {Schmidt}}, \bibinfo {author} {\bibfnamefont {P.}~\bibnamefont {Loubeyre}},\ and\ \bibinfo {author} {\bibfnamefont {J.-F. m.~c.}\ \bibnamefont {Roch}},\ }\bibfield  {title} {\bibinfo {title} {Enabling quantum sensing under extreme pressure: Nitrogen-vacancy magnetometry up to 130 gpa},\ }\href {https://doi.org/10.1103/PhysRevB.107.L220102} {\bibfield  {journal} {\bibinfo  {journal} {Phys. Rev. B}\ }\textbf {\bibinfo {volume} {107}},\ \bibinfo {pages} {L220102} (\bibinfo {year} {2023})}\BibitemShut {NoStop}%
\bibitem [{\citenamefont {Gruber}\ \emph {et~al.}(1997)\citenamefont {Gruber}, \citenamefont {Dr{\"a}benstedt}, \citenamefont {Tietz}, \citenamefont {Fleury}, \citenamefont {Wrachtrup},\ and\ \citenamefont {Borczyskowski}}]{Gruber1997Scanning}%
  \BibitemOpen
  \bibfield  {author} {\bibinfo {author} {\bibfnamefont {A.}~\bibnamefont {Gruber}}, \bibinfo {author} {\bibfnamefont {A.}~\bibnamefont {Dr{\"a}benstedt}}, \bibinfo {author} {\bibfnamefont {C.}~\bibnamefont {Tietz}}, \bibinfo {author} {\bibfnamefont {L.}~\bibnamefont {Fleury}}, \bibinfo {author} {\bibfnamefont {J.}~\bibnamefont {Wrachtrup}},\ and\ \bibinfo {author} {\bibfnamefont {C.~v.}\ \bibnamefont {Borczyskowski}},\ }\bibfield  {title} {\bibinfo {title} {Scanning confocal optical microscopy and magnetic resonance on single defect centers},\ }\href {https://doi.org/10.1126/science.276.5321.2012} {\bibfield  {journal} {\bibinfo  {journal} {Science}\ }\textbf {\bibinfo {volume} {276}},\ \bibinfo {pages} {2012} (\bibinfo {year} {1997})},\ \Eprint {https://arxiv.org/abs/https://science.sciencemag.org/content/276/5321/2012.full.pdf} {https://science.sciencemag.org/content/276/5321/2012.full.pdf} \BibitemShut {NoStop}%
\bibitem [{\citenamefont {Schilling}\ \emph {et~al.}(1990)\citenamefont {Schilling}, \citenamefont {Bernasconi}, \citenamefont {Ott},\ and\ \citenamefont {Hulliger}}]{Schilling1990Specific}%
  \BibitemOpen
  \bibfield  {author} {\bibinfo {author} {\bibfnamefont {A.}~\bibnamefont {Schilling}}, \bibinfo {author} {\bibfnamefont {A.}~\bibnamefont {Bernasconi}}, \bibinfo {author} {\bibfnamefont {H.}~\bibnamefont {Ott}},\ and\ \bibinfo {author} {\bibfnamefont {F.}~\bibnamefont {Hulliger}},\ }\bibfield  {title} {\bibinfo {title} {Specific-heat, resistivity and magnetization study on polycrystalline yba2cu4o8},\ }\href {https://doi.org/https://doi.org/10.1016/0921-4534(90)90180-M} {\bibfield  {journal} {\bibinfo  {journal} {Physica C: Superconductivity}\ }\textbf {\bibinfo {volume} {169}},\ \bibinfo {pages} {237} (\bibinfo {year} {1990})}\BibitemShut {NoStop}%
\bibitem [{\citenamefont {Bucher}\ \emph {et~al.}(1990)\citenamefont {Bucher}, \citenamefont {Karpinski}, \citenamefont {Kaldis},\ and\ \citenamefont {Wachter}}]{Bucher1990Anisotropic}%
  \BibitemOpen
  \bibfield  {author} {\bibinfo {author} {\bibfnamefont {B.}~\bibnamefont {Bucher}}, \bibinfo {author} {\bibfnamefont {J.}~\bibnamefont {Karpinski}}, \bibinfo {author} {\bibfnamefont {E.}~\bibnamefont {Kaldis}},\ and\ \bibinfo {author} {\bibfnamefont {P.}~\bibnamefont {Wachter}},\ }\bibfield  {title} {\bibinfo {title} {Anisotropic behavior in untwinned yba2cu4o8: Optical, magnetic and transport measurements at high pressure},\ }\href {https://doi.org/https://doi.org/10.1016/0921-4534(90)90350-N} {\bibfield  {journal} {\bibinfo  {journal} {Physica C: Superconductivity}\ }\textbf {\bibinfo {volume} {167}},\ \bibinfo {pages} {324} (\bibinfo {year} {1990})}\BibitemShut {NoStop}%
\bibitem [{\citenamefont {Varshney}\ \emph {et~al.}(1996)\citenamefont {Varshney}, \citenamefont {Singh},\ and\ \citenamefont {Shah}}]{Varshney1996Coherence}%
  \BibitemOpen
  \bibfield  {author} {\bibinfo {author} {\bibfnamefont {D.}~\bibnamefont {Varshney}}, \bibinfo {author} {\bibfnamefont {R.~K.}\ \bibnamefont {Singh}},\ and\ \bibinfo {author} {\bibfnamefont {S.}~\bibnamefont {Shah}},\ }\bibfield  {title} {\bibinfo {title} {Coherence lengths and magnetic penetration depths in yba2cu3o7 and yba2cu4o8 superconductors},\ }\href {https://doi.org/10.1007/BF00728246} {\bibfield  {journal} {\bibinfo  {journal} {Journal of Superconductivity}\ }\textbf {\bibinfo {volume} {9}},\ \bibinfo {pages} {629} (\bibinfo {year} {1996})}\BibitemShut {NoStop}%
\bibitem [{\citenamefont {Panagopoulos}\ \emph {et~al.}(1999)\citenamefont {Panagopoulos}, \citenamefont {Tallon},\ and\ \citenamefont {Xiang}}]{Panagopoulos1999Effects}%
  \BibitemOpen
  \bibfield  {author} {\bibinfo {author} {\bibfnamefont {C.}~\bibnamefont {Panagopoulos}}, \bibinfo {author} {\bibfnamefont {J.~L.}\ \bibnamefont {Tallon}},\ and\ \bibinfo {author} {\bibfnamefont {T.}~\bibnamefont {Xiang}},\ }\bibfield  {title} {\bibinfo {title} {Effects of the cu-o chains on the anisotropic penetration depth of ${\mathrm{yba}}_{2}{\mathrm{cu}}_{4}{\mathrm{o}}_{8}$},\ }\href {https://doi.org/10.1103/PhysRevB.59.R6635} {\bibfield  {journal} {\bibinfo  {journal} {Phys. Rev. B}\ }\textbf {\bibinfo {volume} {59}},\ \bibinfo {pages} {R6635} (\bibinfo {year} {1999})}\BibitemShut {NoStop}%
\bibitem [{\citenamefont {Basov}\ \emph {et~al.}(1994)\citenamefont {Basov}, \citenamefont {Timusk}, \citenamefont {Dabrowski},\ and\ \citenamefont {Jorgensen}}]{Basov1994Caxis}%
  \BibitemOpen
  \bibfield  {author} {\bibinfo {author} {\bibfnamefont {D.~N.}\ \bibnamefont {Basov}}, \bibinfo {author} {\bibfnamefont {T.}~\bibnamefont {Timusk}}, \bibinfo {author} {\bibfnamefont {B.}~\bibnamefont {Dabrowski}},\ and\ \bibinfo {author} {\bibfnamefont {J.~D.}\ \bibnamefont {Jorgensen}},\ }\bibfield  {title} {\bibinfo {title} {c-axis response of ${\mathrm{yba}}_{2}$${\mathrm{cu}}_{4}$${\mathrm{o}}_{8}$: A pseudogap and possibility of josephson coupling of ${\mathrm{cuo}}_{2}$ planes},\ }\href {https://doi.org/10.1103/PhysRevB.50.3511} {\bibfield  {journal} {\bibinfo  {journal} {Phys. Rev. B}\ }\textbf {\bibinfo {volume} {50}},\ \bibinfo {pages} {3511} (\bibinfo {year} {1994})}\BibitemShut {NoStop}%
\bibitem [{\citenamefont {Basov}\ \emph {et~al.}(1995)\citenamefont {Basov}, \citenamefont {Liang}, \citenamefont {Bonn}, \citenamefont {Hardy}, \citenamefont {Dabrowski}, \citenamefont {Quijada}, \citenamefont {Tanner}, \citenamefont {Rice}, \citenamefont {Ginsberg},\ and\ \citenamefont {Timusk}}]{Basov1995InPlane}%
  \BibitemOpen
  \bibfield  {author} {\bibinfo {author} {\bibfnamefont {D.~N.}\ \bibnamefont {Basov}}, \bibinfo {author} {\bibfnamefont {R.}~\bibnamefont {Liang}}, \bibinfo {author} {\bibfnamefont {D.~A.}\ \bibnamefont {Bonn}}, \bibinfo {author} {\bibfnamefont {W.~N.}\ \bibnamefont {Hardy}}, \bibinfo {author} {\bibfnamefont {B.}~\bibnamefont {Dabrowski}}, \bibinfo {author} {\bibfnamefont {M.}~\bibnamefont {Quijada}}, \bibinfo {author} {\bibfnamefont {D.~B.}\ \bibnamefont {Tanner}}, \bibinfo {author} {\bibfnamefont {J.~P.}\ \bibnamefont {Rice}}, \bibinfo {author} {\bibfnamefont {D.~M.}\ \bibnamefont {Ginsberg}},\ and\ \bibinfo {author} {\bibfnamefont {T.}~\bibnamefont {Timusk}},\ }\bibfield  {title} {\bibinfo {title} {In-plane anisotropy of the penetration depth in $\mathrm{Y}{\mathrm{ba}}_{2}{\mathrm{cu}}_{3}{\mathrm{o}}_{7\ensuremath{-}x}$ and y${\mathrm{ba}}_{2}$${\mathrm{cu}}_{4}$${\mathrm{o}}_{8}$ superconductors},\ }\href {https://doi.org/10.1103/PhysRevLett.74.598} {\bibfield  {journal} {\bibinfo  {journal} {Phys.
  Rev. Lett.}\ }\textbf {\bibinfo {volume} {74}},\ \bibinfo {pages} {598} (\bibinfo {year} {1995})}\BibitemShut {NoStop}%
\bibitem [{\citenamefont {Grissonnanche}\ \emph {et~al.}(2014)\citenamefont {Grissonnanche}, \citenamefont {Cyr-Choini{\`e}re}, \citenamefont {Lalibert{\'e}}, \citenamefont {Ren{\'e}~de Cotret}, \citenamefont {Juneau-Fecteau}, \citenamefont {Dufour-Beaus{\'e}jour}, \citenamefont {Delage}, \citenamefont {LeBoeuf}, \citenamefont {Chang}, \citenamefont {Ramshaw}, \citenamefont {Bonn}, \citenamefont {Hardy}, \citenamefont {Liang}, \citenamefont {Adachi}, \citenamefont {Hussey}, \citenamefont {Vignolle}, \citenamefont {Proust}, \citenamefont {Sutherland}, \citenamefont {Kr{\"a}mer}, \citenamefont {Park}, \citenamefont {Graf}, \citenamefont {Doiron-Leyraud},\ and\ \citenamefont {Taillefer}}]{Grissonnanche2014Direct}%
  \BibitemOpen
  \bibfield  {author} {\bibinfo {author} {\bibfnamefont {G.}~\bibnamefont {Grissonnanche}}, \bibinfo {author} {\bibfnamefont {O.}~\bibnamefont {Cyr-Choini{\`e}re}}, \bibinfo {author} {\bibfnamefont {F.}~\bibnamefont {Lalibert{\'e}}}, \bibinfo {author} {\bibfnamefont {S.}~\bibnamefont {Ren{\'e}~de Cotret}}, \bibinfo {author} {\bibfnamefont {A.}~\bibnamefont {Juneau-Fecteau}}, \bibinfo {author} {\bibfnamefont {S.}~\bibnamefont {Dufour-Beaus{\'e}jour}}, \bibinfo {author} {\bibfnamefont {M.-{\`E}.}\ \bibnamefont {Delage}}, \bibinfo {author} {\bibfnamefont {D.}~\bibnamefont {LeBoeuf}}, \bibinfo {author} {\bibfnamefont {J.}~\bibnamefont {Chang}}, \bibinfo {author} {\bibfnamefont {B.~J.}\ \bibnamefont {Ramshaw}}, \bibinfo {author} {\bibfnamefont {D.~A.}\ \bibnamefont {Bonn}}, \bibinfo {author} {\bibfnamefont {W.~N.}\ \bibnamefont {Hardy}}, \bibinfo {author} {\bibfnamefont {R.}~\bibnamefont {Liang}}, \bibinfo {author} {\bibfnamefont {S.}~\bibnamefont {Adachi}}, \bibinfo {author} {\bibfnamefont {N.~E.}\ \bibnamefont
  {Hussey}}, \bibinfo {author} {\bibfnamefont {B.}~\bibnamefont {Vignolle}}, \bibinfo {author} {\bibfnamefont {C.}~\bibnamefont {Proust}}, \bibinfo {author} {\bibfnamefont {M.}~\bibnamefont {Sutherland}}, \bibinfo {author} {\bibfnamefont {S.}~\bibnamefont {Kr{\"a}mer}}, \bibinfo {author} {\bibfnamefont {J.-H.}\ \bibnamefont {Park}}, \bibinfo {author} {\bibfnamefont {D.}~\bibnamefont {Graf}}, \bibinfo {author} {\bibfnamefont {N.}~\bibnamefont {Doiron-Leyraud}},\ and\ \bibinfo {author} {\bibfnamefont {L.}~\bibnamefont {Taillefer}},\ }\bibfield  {title} {\bibinfo {title} {Direct measurement of the upper critical field in cuprate superconductors},\ }\href {https://doi.org/10.1038/ncomms4280} {\bibfield  {journal} {\bibinfo  {journal} {Nature Communications}\ }\textbf {\bibinfo {volume} {5}},\ \bibinfo {pages} {3280} (\bibinfo {year} {2014})}\BibitemShut {NoStop}%
\bibitem [{\citenamefont {Heyen}\ \emph {et~al.}(1991)\citenamefont {Heyen}, \citenamefont {Cardona}, \citenamefont {Karpinski}, \citenamefont {Kaldis},\ and\ \citenamefont {Rusiecki}}]{Heyen1991Two}%
  \BibitemOpen
  \bibfield  {author} {\bibinfo {author} {\bibfnamefont {E.~T.}\ \bibnamefont {Heyen}}, \bibinfo {author} {\bibfnamefont {M.}~\bibnamefont {Cardona}}, \bibinfo {author} {\bibfnamefont {J.}~\bibnamefont {Karpinski}}, \bibinfo {author} {\bibfnamefont {E.}~\bibnamefont {Kaldis}},\ and\ \bibinfo {author} {\bibfnamefont {S.}~\bibnamefont {Rusiecki}},\ }\bibfield  {title} {\bibinfo {title} {Two superconducting gaps and electron-phonon coupling in ${\mathrm{yba}}_{2}$${\mathrm{cu}}_{4}$${\mathrm{o}}_{8}$},\ }\href {https://doi.org/10.1103/PhysRevB.43.12958} {\bibfield  {journal} {\bibinfo  {journal} {Phys. Rev. B}\ }\textbf {\bibinfo {volume} {43}},\ \bibinfo {pages} {12958} (\bibinfo {year} {1991})}\BibitemShut {NoStop}%
\bibitem [{\citenamefont {Khasanov}\ \emph {et~al.}(2023)\citenamefont {Khasanov}, \citenamefont {Shengelaya}, \citenamefont {Conder}, \citenamefont {Karpinski}, \citenamefont {Bussmann-Holder},\ and\ \citenamefont {Keller}}]{Khasanov2023Oxygen}%
  \BibitemOpen
  \bibfield  {author} {\bibinfo {author} {\bibfnamefont {R.}~\bibnamefont {Khasanov}}, \bibinfo {author} {\bibfnamefont {A.}~\bibnamefont {Shengelaya}}, \bibinfo {author} {\bibfnamefont {K.}~\bibnamefont {Conder}}, \bibinfo {author} {\bibfnamefont {J.}~\bibnamefont {Karpinski}}, \bibinfo {author} {\bibfnamefont {A.}~\bibnamefont {Bussmann-Holder}},\ and\ \bibinfo {author} {\bibfnamefont {H.}~\bibnamefont {Keller}},\ }\href@noop {} {\bibinfo {title} {Oxygen isotope effect on the superfluid density within the $d-$wave and $s-$wave pairing channels of yba$_2$cu$_4$o$_8$}} (\bibinfo {year} {2023}),\ \Eprint {https://arxiv.org/abs/2306.08112} {arXiv:2306.08112 [cond-mat.supr-con]} \BibitemShut {NoStop}%
\bibitem [{\citenamefont {Khasanov}\ \emph {et~al.}(2008)\citenamefont {Khasanov}, \citenamefont {Shengelaya}, \citenamefont {Karpinski}, \citenamefont {Bussmann-Holder}, \citenamefont {Keller},\ and\ \citenamefont {M{\"u}ller}}]{Khasanov2008S}%
  \BibitemOpen
  \bibfield  {author} {\bibinfo {author} {\bibfnamefont {R.}~\bibnamefont {Khasanov}}, \bibinfo {author} {\bibfnamefont {A.}~\bibnamefont {Shengelaya}}, \bibinfo {author} {\bibfnamefont {J.}~\bibnamefont {Karpinski}}, \bibinfo {author} {\bibfnamefont {A.}~\bibnamefont {Bussmann-Holder}}, \bibinfo {author} {\bibfnamefont {H.}~\bibnamefont {Keller}},\ and\ \bibinfo {author} {\bibfnamefont {K.~A.}\ \bibnamefont {M{\"u}ller}},\ }\bibfield  {title} {\bibinfo {title} {s-wave symmetry along the c-axis and s+d in-plane superconductivity in bulk yba2cu4o8},\ }\href {https://doi.org/10.1007/s10948-007-0302-z} {\bibfield  {journal} {\bibinfo  {journal} {Journal of Superconductivity and Novel Magnetism}\ }\textbf {\bibinfo {volume} {21}},\ \bibinfo {pages} {81} (\bibinfo {year} {2008})}\BibitemShut {NoStop}%
\bibitem [{\citenamefont {Doherty}\ \emph {et~al.}(2014{\natexlab{a}})\citenamefont {Doherty}, \citenamefont {Acosta}, \citenamefont {Jarmola}, \citenamefont {Barson}, \citenamefont {Manson}, \citenamefont {Budker},\ and\ \citenamefont {Hollenberg}}]{Doherty2014Temperature}%
  \BibitemOpen
  \bibfield  {author} {\bibinfo {author} {\bibfnamefont {M.~W.}\ \bibnamefont {Doherty}}, \bibinfo {author} {\bibfnamefont {V.~M.}\ \bibnamefont {Acosta}}, \bibinfo {author} {\bibfnamefont {A.}~\bibnamefont {Jarmola}}, \bibinfo {author} {\bibfnamefont {M.~S.~J.}\ \bibnamefont {Barson}}, \bibinfo {author} {\bibfnamefont {N.~B.}\ \bibnamefont {Manson}}, \bibinfo {author} {\bibfnamefont {D.}~\bibnamefont {Budker}},\ and\ \bibinfo {author} {\bibfnamefont {L.~C.~L.}\ \bibnamefont {Hollenberg}},\ }\bibfield  {title} {\bibinfo {title} {Temperature shifts of the resonances of the ${\mathrm{nv}}^{\ensuremath{-}}$ center in diamond},\ }\href {https://doi.org/10.1103/PhysRevB.90.041201} {\bibfield  {journal} {\bibinfo  {journal} {Phys. Rev. B}\ }\textbf {\bibinfo {volume} {90}},\ \bibinfo {pages} {041201} (\bibinfo {year} {2014}{\natexlab{a}})}\BibitemShut {NoStop}%
\bibitem [{\citenamefont {Chen}\ \emph {et~al.}(2011)\citenamefont {Chen}, \citenamefont {Dong}, \citenamefont {Sun}, \citenamefont {Zou}, \citenamefont {Cui}, \citenamefont {Han},\ and\ \citenamefont {Guo}}]{Chen2011Temperature}%
  \BibitemOpen
  \bibfield  {author} {\bibinfo {author} {\bibfnamefont {X.-D.}\ \bibnamefont {Chen}}, \bibinfo {author} {\bibfnamefont {C.-H.}\ \bibnamefont {Dong}}, \bibinfo {author} {\bibfnamefont {F.-W.}\ \bibnamefont {Sun}}, \bibinfo {author} {\bibfnamefont {C.-L.}\ \bibnamefont {Zou}}, \bibinfo {author} {\bibfnamefont {J.-M.}\ \bibnamefont {Cui}}, \bibinfo {author} {\bibfnamefont {Z.-F.}\ \bibnamefont {Han}},\ and\ \bibinfo {author} {\bibfnamefont {G.-C.}\ \bibnamefont {Guo}},\ }\bibfield  {title} {\bibinfo {title} {Temperature dependent energy level shifts of nitrogen-vacancy centers in diamond},\ }\href {https://doi.org/10.1063/1.3652910} {\bibfield  {journal} {\bibinfo  {journal} {Applied Physics Letters}\ }\textbf {\bibinfo {volume} {99}},\ \bibinfo {pages} {161903} (\bibinfo {year} {2011})},\ \Eprint {https://arxiv.org/abs/https://doi.org/10.1063/1.3652910} {https://doi.org/10.1063/1.3652910} \BibitemShut {NoStop}%
\bibitem [{\citenamefont {Acosta}\ \emph {et~al.}(2010)\citenamefont {Acosta}, \citenamefont {Bauch}, \citenamefont {Ledbetter}, \citenamefont {Waxman}, \citenamefont {Bouchard},\ and\ \citenamefont {Budker}}]{Acosta2010Temperature}%
  \BibitemOpen
  \bibfield  {author} {\bibinfo {author} {\bibfnamefont {V.~M.}\ \bibnamefont {Acosta}}, \bibinfo {author} {\bibfnamefont {E.}~\bibnamefont {Bauch}}, \bibinfo {author} {\bibfnamefont {M.~P.}\ \bibnamefont {Ledbetter}}, \bibinfo {author} {\bibfnamefont {A.}~\bibnamefont {Waxman}}, \bibinfo {author} {\bibfnamefont {L.-S.}\ \bibnamefont {Bouchard}},\ and\ \bibinfo {author} {\bibfnamefont {D.}~\bibnamefont {Budker}},\ }\bibfield  {title} {\bibinfo {title} {Temperature dependence of the nitrogen-vacancy magnetic resonance in diamond},\ }\href {https://doi.org/10.1103/PhysRevLett.104.070801} {\bibfield  {journal} {\bibinfo  {journal} {Phys. Rev. Lett.}\ }\textbf {\bibinfo {volume} {104}},\ \bibinfo {pages} {070801} (\bibinfo {year} {2010})}\BibitemShut {NoStop}%
\bibitem [{\citenamefont {Doherty}\ \emph {et~al.}(2014{\natexlab{b}})\citenamefont {Doherty}, \citenamefont {Struzhkin}, \citenamefont {Simpson}, \citenamefont {McGuinness}, \citenamefont {Meng}, \citenamefont {Stacey}, \citenamefont {Karle}, \citenamefont {Hemley}, \citenamefont {Manson}, \citenamefont {Hollenberg},\ and\ \citenamefont {Prawer}}]{Doherty2014Electronic}%
  \BibitemOpen
  \bibfield  {author} {\bibinfo {author} {\bibfnamefont {M.~W.}\ \bibnamefont {Doherty}}, \bibinfo {author} {\bibfnamefont {V.~V.}\ \bibnamefont {Struzhkin}}, \bibinfo {author} {\bibfnamefont {D.~A.}\ \bibnamefont {Simpson}}, \bibinfo {author} {\bibfnamefont {L.~P.}\ \bibnamefont {McGuinness}}, \bibinfo {author} {\bibfnamefont {Y.}~\bibnamefont {Meng}}, \bibinfo {author} {\bibfnamefont {A.}~\bibnamefont {Stacey}}, \bibinfo {author} {\bibfnamefont {T.~J.}\ \bibnamefont {Karle}}, \bibinfo {author} {\bibfnamefont {R.~J.}\ \bibnamefont {Hemley}}, \bibinfo {author} {\bibfnamefont {N.~B.}\ \bibnamefont {Manson}}, \bibinfo {author} {\bibfnamefont {L.~C.~L.}\ \bibnamefont {Hollenberg}},\ and\ \bibinfo {author} {\bibfnamefont {S.}~\bibnamefont {Prawer}},\ }\bibfield  {title} {\bibinfo {title} {Electronic properties and metrology applications of the diamond ${\mathrm{nv}}^{\ensuremath{-}}$ center under pressure},\ }\href {https://doi.org/10.1103/PhysRevLett.112.047601} {\bibfield  {journal} {\bibinfo  {journal} {Phys.
  Rev. Lett.}\ }\textbf {\bibinfo {volume} {112}},\ \bibinfo {pages} {047601} (\bibinfo {year} {2014}{\natexlab{b}})}\BibitemShut {NoStop}%
\bibitem [{\citenamefont {Steele}\ \emph {et~al.}(2017)\citenamefont {Steele}, \citenamefont {Lawson}, \citenamefont {Onyszczak}, \citenamefont {Bush}, \citenamefont {Mei}, \citenamefont {Dioguardi}, \citenamefont {King}, \citenamefont {Parker}, \citenamefont {Pines}, \citenamefont {Weir}, \citenamefont {Evans}, \citenamefont {Visbeck}, \citenamefont {Vohra},\ and\ \citenamefont {Curro}}]{Steele2017Optically}%
  \BibitemOpen
  \bibfield  {author} {\bibinfo {author} {\bibfnamefont {L.~G.}\ \bibnamefont {Steele}}, \bibinfo {author} {\bibfnamefont {M.}~\bibnamefont {Lawson}}, \bibinfo {author} {\bibfnamefont {M.}~\bibnamefont {Onyszczak}}, \bibinfo {author} {\bibfnamefont {B.~T.}\ \bibnamefont {Bush}}, \bibinfo {author} {\bibfnamefont {Z.}~\bibnamefont {Mei}}, \bibinfo {author} {\bibfnamefont {A.~P.}\ \bibnamefont {Dioguardi}}, \bibinfo {author} {\bibfnamefont {J.}~\bibnamefont {King}}, \bibinfo {author} {\bibfnamefont {A.}~\bibnamefont {Parker}}, \bibinfo {author} {\bibfnamefont {A.}~\bibnamefont {Pines}}, \bibinfo {author} {\bibfnamefont {S.~T.}\ \bibnamefont {Weir}}, \bibinfo {author} {\bibfnamefont {W.}~\bibnamefont {Evans}}, \bibinfo {author} {\bibfnamefont {K.}~\bibnamefont {Visbeck}}, \bibinfo {author} {\bibfnamefont {Y.~K.}\ \bibnamefont {Vohra}},\ and\ \bibinfo {author} {\bibfnamefont {N.~J.}\ \bibnamefont {Curro}},\ }\bibfield  {title} {\bibinfo {title} {Optically detected magnetic resonance of nitrogen vacancies in a
  diamond anvil cell using designer diamond anvils},\ }\href {https://doi.org/10.1063/1.5004153} {\bibfield  {journal} {\bibinfo  {journal} {Applied Physics Letters}\ }\textbf {\bibinfo {volume} {111}},\ \bibinfo {pages} {221903} (\bibinfo {year} {2017})},\ \Eprint {https://arxiv.org/abs/https://doi.org/10.1063/1.5004153} {https://doi.org/10.1063/1.5004153} \BibitemShut {NoStop}%
\bibitem [{\citenamefont {Barson}\ \emph {et~al.}(2017)\citenamefont {Barson}, \citenamefont {Peddibhotla}, \citenamefont {Ovartchaiyapong}, \citenamefont {Ganesan}, \citenamefont {Taylor}, \citenamefont {Gebert}, \citenamefont {Mielens}, \citenamefont {Koslowski}, \citenamefont {Simpson}, \citenamefont {McGuinness}, \citenamefont {McCallum}, \citenamefont {Prawer}, \citenamefont {Onoda}, \citenamefont {Ohshima}, \citenamefont {Bleszynski~Jayich}, \citenamefont {Jelezko}, \citenamefont {Manson},\ and\ \citenamefont {Doherty}}]{Barson2017Nanomechanical}%
  \BibitemOpen
  \bibfield  {author} {\bibinfo {author} {\bibfnamefont {M.~S.~J.}\ \bibnamefont {Barson}}, \bibinfo {author} {\bibfnamefont {P.}~\bibnamefont {Peddibhotla}}, \bibinfo {author} {\bibfnamefont {P.}~\bibnamefont {Ovartchaiyapong}}, \bibinfo {author} {\bibfnamefont {K.}~\bibnamefont {Ganesan}}, \bibinfo {author} {\bibfnamefont {R.~L.}\ \bibnamefont {Taylor}}, \bibinfo {author} {\bibfnamefont {M.}~\bibnamefont {Gebert}}, \bibinfo {author} {\bibfnamefont {Z.}~\bibnamefont {Mielens}}, \bibinfo {author} {\bibfnamefont {B.}~\bibnamefont {Koslowski}}, \bibinfo {author} {\bibfnamefont {D.~A.}\ \bibnamefont {Simpson}}, \bibinfo {author} {\bibfnamefont {L.~P.}\ \bibnamefont {McGuinness}}, \bibinfo {author} {\bibfnamefont {J.}~\bibnamefont {McCallum}}, \bibinfo {author} {\bibfnamefont {S.}~\bibnamefont {Prawer}}, \bibinfo {author} {\bibfnamefont {S.}~\bibnamefont {Onoda}}, \bibinfo {author} {\bibfnamefont {T.}~\bibnamefont {Ohshima}}, \bibinfo {author} {\bibfnamefont {A.~C.}\ \bibnamefont {Bleszynski~Jayich}}, \bibinfo
  {author} {\bibfnamefont {F.}~\bibnamefont {Jelezko}}, \bibinfo {author} {\bibfnamefont {N.~B.}\ \bibnamefont {Manson}},\ and\ \bibinfo {author} {\bibfnamefont {M.~W.}\ \bibnamefont {Doherty}},\ }\bibfield  {title} {\bibinfo {title} {Nanomechanical sensing using spins in diamond},\ }\href {https://doi.org/10.1021/acs.nanolett.6b04544} {\bibfield  {journal} {\bibinfo  {journal} {Nano Letters}\ }\textbf {\bibinfo {volume} {17}},\ \bibinfo {pages} {1496} (\bibinfo {year} {2017})},\ \bibinfo {note} {pMID: 28146361},\ \Eprint {https://arxiv.org/abs/https://doi.org/10.1021/acs.nanolett.6b04544} {https://doi.org/10.1021/acs.nanolett.6b04544} \BibitemShut {NoStop}%
\bibitem [{\citenamefont {Mittiga}\ \emph {et~al.}(2018)\citenamefont {Mittiga}, \citenamefont {Hsieh}, \citenamefont {Zu}, \citenamefont {Kobrin}, \citenamefont {Machado}, \citenamefont {Bhattacharyya}, \citenamefont {Rui}, \citenamefont {Jarmola}, \citenamefont {Choi}, \citenamefont {Budker},\ and\ \citenamefont {Yao}}]{Mittiga2018Imaging}%
  \BibitemOpen
  \bibfield  {author} {\bibinfo {author} {\bibfnamefont {T.}~\bibnamefont {Mittiga}}, \bibinfo {author} {\bibfnamefont {S.}~\bibnamefont {Hsieh}}, \bibinfo {author} {\bibfnamefont {C.}~\bibnamefont {Zu}}, \bibinfo {author} {\bibfnamefont {B.}~\bibnamefont {Kobrin}}, \bibinfo {author} {\bibfnamefont {F.}~\bibnamefont {Machado}}, \bibinfo {author} {\bibfnamefont {P.}~\bibnamefont {Bhattacharyya}}, \bibinfo {author} {\bibfnamefont {N.~Z.}\ \bibnamefont {Rui}}, \bibinfo {author} {\bibfnamefont {A.}~\bibnamefont {Jarmola}}, \bibinfo {author} {\bibfnamefont {S.}~\bibnamefont {Choi}}, \bibinfo {author} {\bibfnamefont {D.}~\bibnamefont {Budker}},\ and\ \bibinfo {author} {\bibfnamefont {N.~Y.}\ \bibnamefont {Yao}},\ }\bibfield  {title} {\bibinfo {title} {Imaging the local charge environment of nitrogen-vacancy centers in diamond},\ }\href {https://doi.org/10.1103/PhysRevLett.121.246402} {\bibfield  {journal} {\bibinfo  {journal} {Phys. Rev. Lett.}\ }\textbf {\bibinfo {volume} {121}},\ \bibinfo {pages} {246402}
  (\bibinfo {year} {2018})}\BibitemShut {NoStop}%
\bibitem [{\citenamefont {Ho}\ \emph {et~al.}(2021{\natexlab{c}})\citenamefont {Ho}, \citenamefont {Leung}, \citenamefont {Pang}, \citenamefont {Wong}, \citenamefont {Ng},\ and\ \citenamefont {Yang}}]{Ho2021In}%
  \BibitemOpen
  \bibfield  {author} {\bibinfo {author} {\bibfnamefont {K.~O.}\ \bibnamefont {Ho}}, \bibinfo {author} {\bibfnamefont {M.~Y.}\ \bibnamefont {Leung}}, \bibinfo {author} {\bibfnamefont {Y.~Y.}\ \bibnamefont {Pang}}, \bibinfo {author} {\bibfnamefont {K.~C.}\ \bibnamefont {Wong}}, \bibinfo {author} {\bibfnamefont {P.~H.}\ \bibnamefont {Ng}},\ and\ \bibinfo {author} {\bibfnamefont {S.}~\bibnamefont {Yang}},\ }\bibfield  {title} {\bibinfo {title} {In situ studies of stress environment in amorphous solids using negatively charged nitrogen vacancy (nv–) centers in nanodiamond},\ }\href {https://doi.org/10.1021/acsapm.0c00964} {\bibfield  {journal} {\bibinfo  {journal} {ACS Applied Polymer Materials}\ }\textbf {\bibinfo {volume} {3}},\ \bibinfo {pages} {162} (\bibinfo {year} {2021}{\natexlab{c}})},\ \Eprint {https://arxiv.org/abs/https://doi.org/10.1021/acsapm.0c00964} {https://doi.org/10.1021/acsapm.0c00964} \BibitemShut {NoStop}%
\bibitem [{\citenamefont {Shin}\ \emph {et~al.}(2013)\citenamefont {Shin}, \citenamefont {Avalos}, \citenamefont {Butler}, \citenamefont {Wang}, \citenamefont {Seltzer}, \citenamefont {Liu}, \citenamefont {Pines},\ and\ \citenamefont {Bajaj}}]{Shin2013Suppression}%
  \BibitemOpen
  \bibfield  {author} {\bibinfo {author} {\bibfnamefont {C.~S.}\ \bibnamefont {Shin}}, \bibinfo {author} {\bibfnamefont {C.~E.}\ \bibnamefont {Avalos}}, \bibinfo {author} {\bibfnamefont {M.~C.}\ \bibnamefont {Butler}}, \bibinfo {author} {\bibfnamefont {H.-J.}\ \bibnamefont {Wang}}, \bibinfo {author} {\bibfnamefont {S.~J.}\ \bibnamefont {Seltzer}}, \bibinfo {author} {\bibfnamefont {R.-B.}\ \bibnamefont {Liu}}, \bibinfo {author} {\bibfnamefont {A.}~\bibnamefont {Pines}},\ and\ \bibinfo {author} {\bibfnamefont {V.~S.}\ \bibnamefont {Bajaj}},\ }\bibfield  {title} {\bibinfo {title} {Suppression of electron spin decoherence of the diamond nv center by a transverse magnetic field},\ }\href {https://doi.org/10.1103/PhysRevB.88.161412} {\bibfield  {journal} {\bibinfo  {journal} {Phys. Rev. B}\ }\textbf {\bibinfo {volume} {88}},\ \bibinfo {pages} {161412} (\bibinfo {year} {2013})}\BibitemShut {NoStop}%
\bibitem [{\citenamefont {Doherty}\ \emph {et~al.}(2014{\natexlab{c}})\citenamefont {Doherty}, \citenamefont {Michl}, \citenamefont {Dolde}, \citenamefont {Jakobi}, \citenamefont {Neumann}, \citenamefont {Manson},\ and\ \citenamefont {Wrachtrup}}]{Doherty2014Measuring}%
  \BibitemOpen
  \bibfield  {author} {\bibinfo {author} {\bibfnamefont {M.~W.}\ \bibnamefont {Doherty}}, \bibinfo {author} {\bibfnamefont {J.}~\bibnamefont {Michl}}, \bibinfo {author} {\bibfnamefont {F.}~\bibnamefont {Dolde}}, \bibinfo {author} {\bibfnamefont {I.}~\bibnamefont {Jakobi}}, \bibinfo {author} {\bibfnamefont {P.}~\bibnamefont {Neumann}}, \bibinfo {author} {\bibfnamefont {N.~B.}\ \bibnamefont {Manson}},\ and\ \bibinfo {author} {\bibfnamefont {J.}~\bibnamefont {Wrachtrup}},\ }\bibfield  {title} {\bibinfo {title} {Measuring the defect structure orientation of a single nv$^{-}$ centre in diamond},\ }\href {https://doi.org/10.1088/1367-2630/16/6/063067} {\bibfield  {journal} {\bibinfo  {journal} {New Journal of Physics}\ }\textbf {\bibinfo {volume} {16}},\ \bibinfo {pages} {063067} (\bibinfo {year} {2014}{\natexlab{c}})}\BibitemShut {NoStop}%
\bibitem [{SI()}]{SI}%
  \BibitemOpen
  \href@noop {} {}\bibinfo {note} {See Supplemental Material for the data analysis and experimental details.}\BibitemShut {Stop}%
\bibitem [{\citenamefont {Krusin-Elbaum}\ \emph {et~al.}(1989)\citenamefont {Krusin-Elbaum}, \citenamefont {Malozemoff}, \citenamefont {Yeshurun}, \citenamefont {Cronemeyer},\ and\ \citenamefont {Holtzberg}}]{Krusin-Elbaum1989Temperature}%
  \BibitemOpen
  \bibfield  {author} {\bibinfo {author} {\bibfnamefont {L.}~\bibnamefont {Krusin-Elbaum}}, \bibinfo {author} {\bibfnamefont {A.~P.}\ \bibnamefont {Malozemoff}}, \bibinfo {author} {\bibfnamefont {Y.}~\bibnamefont {Yeshurun}}, \bibinfo {author} {\bibfnamefont {D.~C.}\ \bibnamefont {Cronemeyer}},\ and\ \bibinfo {author} {\bibfnamefont {F.}~\bibnamefont {Holtzberg}},\ }\bibfield  {title} {\bibinfo {title} {Temperature dependence of lower critical fields in y-ba-cu-o crystals},\ }\href {https://doi.org/10.1103/PhysRevB.39.2936} {\bibfield  {journal} {\bibinfo  {journal} {Phys. Rev. B}\ }\textbf {\bibinfo {volume} {39}},\ \bibinfo {pages} {2936} (\bibinfo {year} {1989})}\BibitemShut {NoStop}%
\bibitem [{\citenamefont {Bean}(1962)}]{Bean1962Magnetization}%
  \BibitemOpen
  \bibfield  {author} {\bibinfo {author} {\bibfnamefont {C.~P.}\ \bibnamefont {Bean}},\ }\bibfield  {title} {\bibinfo {title} {Magnetization of hard superconductors},\ }\href {https://doi.org/10.1103/PhysRevLett.8.250} {\bibfield  {journal} {\bibinfo  {journal} {Phys. Rev. Lett.}\ }\textbf {\bibinfo {volume} {8}},\ \bibinfo {pages} {250} (\bibinfo {year} {1962})}\BibitemShut {NoStop}%
\bibitem [{\citenamefont {BEAN}(1964)}]{Bean1964Magnetization}%
  \BibitemOpen
  \bibfield  {author} {\bibinfo {author} {\bibfnamefont {C.~P.}\ \bibnamefont {BEAN}},\ }\bibfield  {title} {\bibinfo {title} {Magnetization of high-field superconductors},\ }\href {https://doi.org/10.1103/RevModPhys.36.31} {\bibfield  {journal} {\bibinfo  {journal} {Rev. Mod. Phys.}\ }\textbf {\bibinfo {volume} {36}},\ \bibinfo {pages} {31} (\bibinfo {year} {1964})}\BibitemShut {NoStop}%
\bibitem [{\citenamefont {Krusin‐Elbaum}\ \emph {et~al.}(1990)\citenamefont {Krusin‐Elbaum}, \citenamefont {Malozemoff}, \citenamefont {Cronemeyer}, \citenamefont {Holtzberg}, \citenamefont {Clem},\ and\ \citenamefont {Hao}}]{Krusin-Elbaum1990New}%
  \BibitemOpen
  \bibfield  {author} {\bibinfo {author} {\bibfnamefont {L.}~\bibnamefont {Krusin‐Elbaum}}, \bibinfo {author} {\bibfnamefont {A.~P.}\ \bibnamefont {Malozemoff}}, \bibinfo {author} {\bibfnamefont {D.~C.}\ \bibnamefont {Cronemeyer}}, \bibinfo {author} {\bibfnamefont {F.}~\bibnamefont {Holtzberg}}, \bibinfo {author} {\bibfnamefont {J.~R.}\ \bibnamefont {Clem}},\ and\ \bibinfo {author} {\bibfnamefont {Z.}~\bibnamefont {Hao}},\ }\bibfield  {title} {\bibinfo {title} {New mechanisms for irreversibility in high‐tc superconductors (invited)},\ }\href {https://doi.org/10.1063/1.344822} {\bibfield  {journal} {\bibinfo  {journal} {Journal of Applied Physics}\ }\textbf {\bibinfo {volume} {67}},\ \bibinfo {pages} {4670} (\bibinfo {year} {1990})},\ \Eprint {https://arxiv.org/abs/https://doi.org/10.1063/1.344822} {https://doi.org/10.1063/1.344822} \BibitemShut {NoStop}%
\bibitem [{\citenamefont {Tomioka}\ \emph {et~al.}(1994)\citenamefont {Tomioka}, \citenamefont {Naito}, \citenamefont {Kishio},\ and\ \citenamefont {Kitazawa}}]{Tomioka1994The}%
  \BibitemOpen
  \bibfield  {author} {\bibinfo {author} {\bibfnamefont {Y.}~\bibnamefont {Tomioka}}, \bibinfo {author} {\bibfnamefont {M.}~\bibnamefont {Naito}}, \bibinfo {author} {\bibfnamefont {K.}~\bibnamefont {Kishio}},\ and\ \bibinfo {author} {\bibfnamefont {K.}~\bibnamefont {Kitazawa}},\ }\bibfield  {title} {\bibinfo {title} {The meissner and shielding effects of high-temperature oxide superconductors},\ }\href {https://doi.org/https://doi.org/10.1016/0921-4534(94)91279-3} {\bibfield  {journal} {\bibinfo  {journal} {Physica C: Superconductivity}\ }\textbf {\bibinfo {volume} {223}},\ \bibinfo {pages} {347} (\bibinfo {year} {1994})}\BibitemShut {NoStop}%
\bibitem [{\citenamefont {Wolf}\ \emph {et~al.}(1993)\citenamefont {Wolf}, \citenamefont {Gleitzmann},\ and\ \citenamefont {Gey}}]{Wolf1993Relaxation}%
  \BibitemOpen
  \bibfield  {author} {\bibinfo {author} {\bibfnamefont {M.}~\bibnamefont {Wolf}}, \bibinfo {author} {\bibfnamefont {J.}~\bibnamefont {Gleitzmann}},\ and\ \bibinfo {author} {\bibfnamefont {W.}~\bibnamefont {Gey}},\ }\bibfield  {title} {\bibinfo {title} {Relaxation-caused suppression of magnetization in superconducting ${\mathrm{yba}}_{2}$${\mathrm{cu}}_{3}$${\mathrm{o}}_{7\mathrm{\ensuremath{-}}\mathrm{\ensuremath{\delta}}}$ single crystals},\ }\href {https://doi.org/10.1103/PhysRevB.47.8381} {\bibfield  {journal} {\bibinfo  {journal} {Phys. Rev. B}\ }\textbf {\bibinfo {volume} {47}},\ \bibinfo {pages} {8381} (\bibinfo {year} {1993})}\BibitemShut {NoStop}%
\bibitem [{\citenamefont {Kumar}\ and\ \citenamefont {Chaddah}(1989)}]{Kumar1989Extension}%
  \BibitemOpen
  \bibfield  {author} {\bibinfo {author} {\bibfnamefont {G.~R.}\ \bibnamefont {Kumar}}\ and\ \bibinfo {author} {\bibfnamefont {P.}~\bibnamefont {Chaddah}},\ }\bibfield  {title} {\bibinfo {title} {Extension of bean's model for high-${T}_{c}$ superconductors},\ }\href {https://doi.org/10.1103/PhysRevB.39.4704} {\bibfield  {journal} {\bibinfo  {journal} {Phys. Rev. B}\ }\textbf {\bibinfo {volume} {39}},\ \bibinfo {pages} {4704} (\bibinfo {year} {1989})}\BibitemShut {NoStop}%
\bibitem [{\citenamefont {Brandt}\ and\ \citenamefont {Indenbom}(1993)}]{Brandt1993TypeII}%
  \BibitemOpen
  \bibfield  {author} {\bibinfo {author} {\bibfnamefont {E.~H.}\ \bibnamefont {Brandt}}\ and\ \bibinfo {author} {\bibfnamefont {M.}~\bibnamefont {Indenbom}},\ }\bibfield  {title} {\bibinfo {title} {Type-ii-superconductor strip with current in a perpendicular magnetic field},\ }\href {https://doi.org/10.1103/PhysRevB.48.12893} {\bibfield  {journal} {\bibinfo  {journal} {Phys. Rev. B}\ }\textbf {\bibinfo {volume} {48}},\ \bibinfo {pages} {12893} (\bibinfo {year} {1993})}\BibitemShut {NoStop}%
\bibitem [{\citenamefont {Brandt}(1999)}]{Brandt1996Geometric}%
  \BibitemOpen
  \bibfield  {author} {\bibinfo {author} {\bibfnamefont {E.~H.}\ \bibnamefont {Brandt}},\ }\bibfield  {title} {\bibinfo {title} {Geometric barrier and current string in type-ii superconductors obtained from continuum electrodynamics},\ }\href {https://doi.org/10.1103/PhysRevB.59.3369} {\bibfield  {journal} {\bibinfo  {journal} {Phys. Rev. B}\ }\textbf {\bibinfo {volume} {59}},\ \bibinfo {pages} {3369} (\bibinfo {year} {1999})}\BibitemShut {NoStop}%
\bibitem [{\citenamefont {Abrikosov}(1957)}]{Abrikosov1957The}%
  \BibitemOpen
  \bibfield  {author} {\bibinfo {author} {\bibfnamefont {A.}~\bibnamefont {Abrikosov}},\ }\bibfield  {title} {\bibinfo {title} {The magnetic properties of superconducting alloys},\ }\href {https://doi.org/https://doi.org/10.1016/0022-3697(57)90083-5} {\bibfield  {journal} {\bibinfo  {journal} {Journal of Physics and Chemistry of Solids}\ }\textbf {\bibinfo {volume} {2}},\ \bibinfo {pages} {199} (\bibinfo {year} {1957})}\BibitemShut {NoStop}%
\bibitem [{\citenamefont {Auslaender}\ \emph {et~al.}(2009)\citenamefont {Auslaender}, \citenamefont {Luan}, \citenamefont {Straver}, \citenamefont {Hoffman}, \citenamefont {Koshnick}, \citenamefont {Zeldov}, \citenamefont {Bonn}, \citenamefont {Liang}, \citenamefont {Hardy},\ and\ \citenamefont {Moler}}]{Auslaender2009Mechanics}%
  \BibitemOpen
  \bibfield  {author} {\bibinfo {author} {\bibfnamefont {O.~M.}\ \bibnamefont {Auslaender}}, \bibinfo {author} {\bibfnamefont {L.}~\bibnamefont {Luan}}, \bibinfo {author} {\bibfnamefont {E.~W.~J.}\ \bibnamefont {Straver}}, \bibinfo {author} {\bibfnamefont {J.~E.}\ \bibnamefont {Hoffman}}, \bibinfo {author} {\bibfnamefont {N.~C.}\ \bibnamefont {Koshnick}}, \bibinfo {author} {\bibfnamefont {E.}~\bibnamefont {Zeldov}}, \bibinfo {author} {\bibfnamefont {D.~A.}\ \bibnamefont {Bonn}}, \bibinfo {author} {\bibfnamefont {R.}~\bibnamefont {Liang}}, \bibinfo {author} {\bibfnamefont {W.~N.}\ \bibnamefont {Hardy}},\ and\ \bibinfo {author} {\bibfnamefont {K.~A.}\ \bibnamefont {Moler}},\ }\bibfield  {title} {\bibinfo {title} {Mechanics of individual isolated vortices in a cuprate superconductor},\ }\href {https://doi.org/10.1038/nphys1127} {\bibfield  {journal} {\bibinfo  {journal} {Nature Physics}\ }\textbf {\bibinfo {volume} {5}},\ \bibinfo {pages} {35} (\bibinfo {year} {2009})}\BibitemShut {NoStop}%
\bibitem [{\citenamefont {Prozorov}\ and\ \citenamefont {Kogan}(2018)}]{Prozorov2018Effective}%
  \BibitemOpen
  \bibfield  {author} {\bibinfo {author} {\bibfnamefont {R.}~\bibnamefont {Prozorov}}\ and\ \bibinfo {author} {\bibfnamefont {V.~G.}\ \bibnamefont {Kogan}},\ }\bibfield  {title} {\bibinfo {title} {Effective demagnetizing factors of diamagnetic samples of various shapes},\ }\href {https://doi.org/10.1103/PhysRevApplied.10.014030} {\bibfield  {journal} {\bibinfo  {journal} {Phys. Rev. Applied}\ }\textbf {\bibinfo {volume} {10}},\ \bibinfo {pages} {014030} (\bibinfo {year} {2018})}\BibitemShut {NoStop}%
\bibitem [{\citenamefont {Gao}\ and\ \citenamefont {O’Brien}(1993)}]{Gao1993A}%
  \BibitemOpen
  \bibfield  {author} {\bibinfo {author} {\bibfnamefont {F.}~\bibnamefont {Gao}}\ and\ \bibinfo {author} {\bibfnamefont {E.~E.}\ \bibnamefont {O’Brien}},\ }\bibfield  {title} {\bibinfo {title} {{A large‐eddy simulation scheme for turbulent reacting flows}},\ }\href {https://doi.org/10.1063/1.858617} {\bibfield  {journal} {\bibinfo  {journal} {Physics of Fluids A: Fluid Dynamics}\ }\textbf {\bibinfo {volume} {5}},\ \bibinfo {pages} {1282} (\bibinfo {year} {1993})},\ \Eprint {https://arxiv.org/abs/https://pubs.aip.org/aip/pof/article-pdf/5/6/1282/12525781/1282\_1\_online.pdf} {https://pubs.aip.org/aip/pof/article-pdf/5/6/1282/12525781/1282\_1\_online.pdf} \BibitemShut {NoStop}%
\bibitem [{\citenamefont {Knikker}\ \emph {et~al.}(2004)\citenamefont {Knikker}, \citenamefont {Veynante},\ and\ \citenamefont {Meneveau}}]{Knikker2004A}%
  \BibitemOpen
  \bibfield  {author} {\bibinfo {author} {\bibfnamefont {R.}~\bibnamefont {Knikker}}, \bibinfo {author} {\bibfnamefont {D.}~\bibnamefont {Veynante}},\ and\ \bibinfo {author} {\bibfnamefont {C.}~\bibnamefont {Meneveau}},\ }\bibfield  {title} {\bibinfo {title} {{A dynamic flame surface density model for large eddy simulation of turbulent premixed combustion}},\ }\href {https://doi.org/10.1063/1.1780549} {\bibfield  {journal} {\bibinfo  {journal} {Physics of Fluids}\ }\textbf {\bibinfo {volume} {16}},\ \bibinfo {pages} {L91} (\bibinfo {year} {2004})},\ \Eprint {https://arxiv.org/abs/https://pubs.aip.org/aip/pof/article-pdf/16/11/L91/12455597/l91\_1\_online.pdf} {https://pubs.aip.org/aip/pof/article-pdf/16/11/L91/12455597/l91\_1\_online.pdf} \BibitemShut {NoStop}%
\bibitem [{\citenamefont {Itatani}\ \emph {et~al.}(2013)\citenamefont {Itatani}, \citenamefont {Okada}, \citenamefont {Uejima}, \citenamefont {Tanaka}, \citenamefont {Ono}, \citenamefont {Miyaji},\ and\ \citenamefont {Takenaka}}]{Itatani2013Intraventricular}%
  \BibitemOpen
  \bibfield  {author} {\bibinfo {author} {\bibfnamefont {K.}~\bibnamefont {Itatani}}, \bibinfo {author} {\bibfnamefont {T.}~\bibnamefont {Okada}}, \bibinfo {author} {\bibfnamefont {T.}~\bibnamefont {Uejima}}, \bibinfo {author} {\bibfnamefont {T.}~\bibnamefont {Tanaka}}, \bibinfo {author} {\bibfnamefont {M.}~\bibnamefont {Ono}}, \bibinfo {author} {\bibfnamefont {K.}~\bibnamefont {Miyaji}},\ and\ \bibinfo {author} {\bibfnamefont {K.}~\bibnamefont {Takenaka}},\ }\bibfield  {title} {\bibinfo {title} {Intraventricular flow velocity vector visualization based on the continuity equation and measurements of vorticity and wall shear stress},\ }\href {https://doi.org/10.7567/JJAP.52.07HF16} {\bibfield  {journal} {\bibinfo  {journal} {Japanese Journal of Applied Physics}\ }\textbf {\bibinfo {volume} {52}},\ \bibinfo {pages} {07HF16} (\bibinfo {year} {2013})}\BibitemShut {NoStop}%
\bibitem [{\citenamefont {Capellera}\ \emph {et~al.}(2021)\citenamefont {Capellera}, \citenamefont {Ianniciello}, \citenamefont {Romanini},\ and\ \citenamefont {Vives}}]{Capellera2021Heat}%
  \BibitemOpen
  \bibfield  {author} {\bibinfo {author} {\bibfnamefont {G.}~\bibnamefont {Capellera}}, \bibinfo {author} {\bibfnamefont {L.}~\bibnamefont {Ianniciello}}, \bibinfo {author} {\bibfnamefont {M.}~\bibnamefont {Romanini}},\ and\ \bibinfo {author} {\bibfnamefont {E.}~\bibnamefont {Vives}},\ }\bibfield  {title} {\bibinfo {title} {{Heat sink avalanche dynamics in elastocaloric Cu–Al–Ni single crystal detected by infrared calorimetry and Gaussian filtering}},\ }\bibfield  {journal} {\bibinfo  {journal} {Applied Physics Letters}\ }\textbf {\bibinfo {volume} {119}},\ \href {https://doi.org/10.1063/5.0066525} {10.1063/5.0066525} (\bibinfo {year} {2021}),\ \bibinfo {note} {151905},\ \Eprint {https://arxiv.org/abs/https://pubs.aip.org/aip/apl/article-pdf/doi/10.1063/5.0066525/14553956/151905\_1\_online.pdf} {https://pubs.aip.org/aip/apl/article-pdf/doi/10.1063/5.0066525/14553956/151905\_1\_online.pdf} \BibitemShut {NoStop}%
\bibitem [{\citenamefont {Gulian}(2020)}]{Gulian2020Exploring}%
  \BibitemOpen
  \bibfield  {author} {\bibinfo {author} {\bibfnamefont {A.}~\bibnamefont {Gulian}},\ }\bibinfo {title} {Exploring superconductivity with comsol via tdgl equations},\ in\ \href {https://doi.org/10.1007/978-3-030-23486-7_2} {\emph {\bibinfo {booktitle} {Shortcut to Superconductivity: Superconducting Electronics via COMSOL Modeling}}}\ (\bibinfo  {publisher} {Springer International Publishing},\ \bibinfo {address} {Cham},\ \bibinfo {year} {2020})\ pp.\ \bibinfo {pages} {51--110}\BibitemShut {NoStop}%
\bibitem [{\citenamefont {Naito}\ \emph {et~al.}(2016)\citenamefont {Naito}, \citenamefont {Yamamoto}, \citenamefont {Konishi}, \citenamefont {Kubo}, \citenamefont {Nakamura},\ and\ \citenamefont {Moyama}}]{Naito2016Critical}%
  \BibitemOpen
  \bibfield  {author} {\bibinfo {author} {\bibfnamefont {T.}~\bibnamefont {Naito}}, \bibinfo {author} {\bibfnamefont {H.}~\bibnamefont {Yamamoto}}, \bibinfo {author} {\bibfnamefont {K.}~\bibnamefont {Konishi}}, \bibinfo {author} {\bibfnamefont {K.}~\bibnamefont {Kubo}}, \bibinfo {author} {\bibfnamefont {T.}~\bibnamefont {Nakamura}},\ and\ \bibinfo {author} {\bibfnamefont {H.}~\bibnamefont {Moyama}},\ }\bibfield  {title} {\bibinfo {title} {Critical current density of superconductors with different fractal dimensions},\ }\href {https://doi.org/10.15761/AMS.1000105} {\bibfield  {journal} {\bibinfo  {journal} {Adv. Mater. Sci}\ }\textbf {\bibinfo {volume} {1}},\ \bibinfo {pages} {15} (\bibinfo {year} {2016})}\BibitemShut {NoStop}%
\bibitem [{\citenamefont {Lombardo}\ \emph {et~al.}(2011)\citenamefont {Lombardo}, \citenamefont {Barzi}, \citenamefont {Turrioni},\ and\ \citenamefont {Zlobin}}]{Lombardo2011Critical}%
  \BibitemOpen
  \bibfield  {author} {\bibinfo {author} {\bibfnamefont {V.}~\bibnamefont {Lombardo}}, \bibinfo {author} {\bibfnamefont {E.}~\bibnamefont {Barzi}}, \bibinfo {author} {\bibfnamefont {D.}~\bibnamefont {Turrioni}},\ and\ \bibinfo {author} {\bibfnamefont {A.~V.}\ \bibnamefont {Zlobin}},\ }\bibfield  {title} {\bibinfo {title} {Critical currents of ${\rm yba}_{2}{\rm cu}_{3}{\rm o}_{7-\delta}$ tapes and ${\rm bi}_{2}{\rm sr}_{2}{\rm cacu}_{2}{\rm o}_{\rm x}$ wires at different temperatures and magnetic fields},\ }\href {https://doi.org/10.1109/TASC.2010.2093865} {\bibfield  {journal} {\bibinfo  {journal} {IEEE Transactions on Applied Superconductivity}\ }\textbf {\bibinfo {volume} {21}},\ \bibinfo {pages} {3247} (\bibinfo {year} {2011})}\BibitemShut {NoStop}%
\bibitem [{\citenamefont {Braccini}\ \emph {et~al.}(2010)\citenamefont {Braccini}, \citenamefont {Xu}, \citenamefont {Jaroszynski}, \citenamefont {Xin}, \citenamefont {Larbalestier}, \citenamefont {Chen}, \citenamefont {Carota}, \citenamefont {Dackow}, \citenamefont {Kesgin}, \citenamefont {Yao}, \citenamefont {Guevara}, \citenamefont {Shi},\ and\ \citenamefont {Selvamanickam}}]{Braccini2011Properties}%
  \BibitemOpen
  \bibfield  {author} {\bibinfo {author} {\bibfnamefont {V.}~\bibnamefont {Braccini}}, \bibinfo {author} {\bibfnamefont {A.}~\bibnamefont {Xu}}, \bibinfo {author} {\bibfnamefont {J.}~\bibnamefont {Jaroszynski}}, \bibinfo {author} {\bibfnamefont {Y.}~\bibnamefont {Xin}}, \bibinfo {author} {\bibfnamefont {D.~C.}\ \bibnamefont {Larbalestier}}, \bibinfo {author} {\bibfnamefont {Y.}~\bibnamefont {Chen}}, \bibinfo {author} {\bibfnamefont {G.}~\bibnamefont {Carota}}, \bibinfo {author} {\bibfnamefont {J.}~\bibnamefont {Dackow}}, \bibinfo {author} {\bibfnamefont {I.}~\bibnamefont {Kesgin}}, \bibinfo {author} {\bibfnamefont {Y.}~\bibnamefont {Yao}}, \bibinfo {author} {\bibfnamefont {A.}~\bibnamefont {Guevara}}, \bibinfo {author} {\bibfnamefont {T.}~\bibnamefont {Shi}},\ and\ \bibinfo {author} {\bibfnamefont {V.}~\bibnamefont {Selvamanickam}},\ }\bibfield  {title} {\bibinfo {title} {Properties of recent ibad–mocvd coated conductors relevant to their high field, low temperature magnet use},\ }\href
  {https://doi.org/10.1088/0953-2048/24/3/035001} {\bibfield  {journal} {\bibinfo  {journal} {Superconductor Science and Technology}\ }\textbf {\bibinfo {volume} {24}},\ \bibinfo {pages} {035001} (\bibinfo {year} {2010})}\BibitemShut {NoStop}%
\bibitem [{\citenamefont {Pol{\'a}k}\ \emph {et~al.}(2002)\citenamefont {Pol{\'a}k}, \citenamefont {Krempask{\'y}}, \citenamefont {Chromik}, \citenamefont {Wehler},\ and\ \citenamefont {Moenter}}]{Polak2002Magnetic}%
  \BibitemOpen
  \bibfield  {author} {\bibinfo {author} {\bibfnamefont {M.}~\bibnamefont {Pol{\'a}k}}, \bibinfo {author} {\bibfnamefont {L.}~\bibnamefont {Krempask{\'y}}}, \bibinfo {author} {\bibfnamefont {{\v{S}}.}~\bibnamefont {Chromik}}, \bibinfo {author} {\bibfnamefont {D.}~\bibnamefont {Wehler}},\ and\ \bibinfo {author} {\bibfnamefont {B.}~\bibnamefont {Moenter}},\ }\bibfield  {title} {\bibinfo {title} {Magnetic field in the vicinity of ybco thin film strip and strip with filamentary structure},\ }\href {https://doi.org/https://doi.org/10.1016/S0921-4534(02)01002-X} {\bibfield  {journal} {\bibinfo  {journal} {Physica C: Superconductivity}\ }\textbf {\bibinfo {volume} {372-376}},\ \bibinfo {pages} {1830} (\bibinfo {year} {2002})}\BibitemShut {NoStop}%
\bibitem [{\citenamefont {Beck}\ \emph {et~al.}(2004)\citenamefont {Beck}, \citenamefont {Leibovitch}, \citenamefont {Milner}, \citenamefont {Gerber},\ and\ \citenamefont {Deutscher}}]{Beck2004Determination}%
  \BibitemOpen
  \bibfield  {author} {\bibinfo {author} {\bibfnamefont {R.}~\bibnamefont {Beck}}, \bibinfo {author} {\bibfnamefont {G.}~\bibnamefont {Leibovitch}}, \bibinfo {author} {\bibfnamefont {A.}~\bibnamefont {Milner}}, \bibinfo {author} {\bibfnamefont {A.}~\bibnamefont {Gerber}},\ and\ \bibinfo {author} {\bibfnamefont {G.}~\bibnamefont {Deutscher}},\ }\bibfield  {title} {\bibinfo {title} {Determination of the critical current density in the d-wave superconductor ybco under applied magnetic fields by nodal tunnelling},\ }\href {https://doi.org/10.1088/0953-2048/17/8/022} {\bibfield  {journal} {\bibinfo  {journal} {Superconductor Science and Technology}\ }\textbf {\bibinfo {volume} {17}},\ \bibinfo {pages} {1069} (\bibinfo {year} {2004})}\BibitemShut {NoStop}%
\bibitem [{\citenamefont {Maertz}\ \emph {et~al.}(2010)\citenamefont {Maertz}, \citenamefont {Wijnheijmer}, \citenamefont {Fuchs}, \citenamefont {Nowakowski},\ and\ \citenamefont {Awschalom}}]{Maertz2010Vector}%
  \BibitemOpen
  \bibfield  {author} {\bibinfo {author} {\bibfnamefont {B.~J.}\ \bibnamefont {Maertz}}, \bibinfo {author} {\bibfnamefont {A.~P.}\ \bibnamefont {Wijnheijmer}}, \bibinfo {author} {\bibfnamefont {G.~D.}\ \bibnamefont {Fuchs}}, \bibinfo {author} {\bibfnamefont {M.~E.}\ \bibnamefont {Nowakowski}},\ and\ \bibinfo {author} {\bibfnamefont {D.~D.}\ \bibnamefont {Awschalom}},\ }\bibfield  {title} {\bibinfo {title} {Vector magnetic field microscopy using nitrogen vacancy centers in diamond},\ }\href {https://doi.org/10.1063/1.3337096} {\bibfield  {journal} {\bibinfo  {journal} {Applied Physics Letters}\ }\textbf {\bibinfo {volume} {96}},\ \bibinfo {pages} {092504} (\bibinfo {year} {2010})},\ \Eprint {https://arxiv.org/abs/https://doi.org/10.1063/1.3337096} {https://doi.org/10.1063/1.3337096} \BibitemShut {NoStop}%
\bibitem [{\citenamefont {Zhang}\ \emph {et~al.}(2022)\citenamefont {Zhang}, \citenamefont {Yin}, \citenamefont {Wang}, \citenamefont {Wang}, \citenamefont {Li}, \citenamefont {Tian},\ and\ \citenamefont {Chen}}]{Zhang2022Single}%
  \BibitemOpen
  \bibfield  {author} {\bibinfo {author} {\bibfnamefont {Z.-D.}\ \bibnamefont {Zhang}}, \bibinfo {author} {\bibfnamefont {S.-Y.}\ \bibnamefont {Yin}}, \bibinfo {author} {\bibfnamefont {L.-C.}\ \bibnamefont {Wang}}, \bibinfo {author} {\bibfnamefont {Y.-D.}\ \bibnamefont {Wang}}, \bibinfo {author} {\bibfnamefont {Y.-F.}\ \bibnamefont {Li}}, \bibinfo {author} {\bibfnamefont {Z.-N.}\ \bibnamefont {Tian}},\ and\ \bibinfo {author} {\bibfnamefont {Q.-D.}\ \bibnamefont {Chen}},\ }\bibfield  {title} {\bibinfo {title} {Single nv centers array preparation and static magnetic field detection},\ }\href {https://doi.org/10.1364/OE.470400} {\bibfield  {journal} {\bibinfo  {journal} {Opt. Express}\ }\textbf {\bibinfo {volume} {30}},\ \bibinfo {pages} {32355} (\bibinfo {year} {2022})}\BibitemShut {NoStop}%
\bibitem [{\citenamefont {Tetienne}\ \emph {et~al.}(2013)\citenamefont {Tetienne}, \citenamefont {Hingant}, \citenamefont {Rondin}, \citenamefont {Cavaill\`es}, \citenamefont {Mayer}, \citenamefont {Dantelle}, \citenamefont {Gacoin}, \citenamefont {Wrachtrup}, \citenamefont {Roch},\ and\ \citenamefont {Jacques}}]{Tetienne2013Spin}%
  \BibitemOpen
  \bibfield  {author} {\bibinfo {author} {\bibfnamefont {J.-P.}\ \bibnamefont {Tetienne}}, \bibinfo {author} {\bibfnamefont {T.}~\bibnamefont {Hingant}}, \bibinfo {author} {\bibfnamefont {L.}~\bibnamefont {Rondin}}, \bibinfo {author} {\bibfnamefont {A.}~\bibnamefont {Cavaill\`es}}, \bibinfo {author} {\bibfnamefont {L.}~\bibnamefont {Mayer}}, \bibinfo {author} {\bibfnamefont {G.}~\bibnamefont {Dantelle}}, \bibinfo {author} {\bibfnamefont {T.}~\bibnamefont {Gacoin}}, \bibinfo {author} {\bibfnamefont {J.}~\bibnamefont {Wrachtrup}}, \bibinfo {author} {\bibfnamefont {J.-F.}\ \bibnamefont {Roch}},\ and\ \bibinfo {author} {\bibfnamefont {V.}~\bibnamefont {Jacques}},\ }\bibfield  {title} {\bibinfo {title} {Spin relaxometry of single nitrogen-vacancy defects in diamond nanocrystals for magnetic noise sensing},\ }\href {https://doi.org/10.1103/PhysRevB.87.235436} {\bibfield  {journal} {\bibinfo  {journal} {Phys. Rev. B}\ }\textbf {\bibinfo {volume} {87}},\ \bibinfo {pages} {235436} (\bibinfo {year} {2013})}\BibitemShut
  {NoStop}%
\bibitem [{\citenamefont {Schmid-Lorch}\ \emph {et~al.}(2015)\citenamefont {Schmid-Lorch}, \citenamefont {Häberle}, \citenamefont {Reinhard}, \citenamefont {Zappe}, \citenamefont {Slota}, \citenamefont {Bogani}, \citenamefont {Finkler},\ and\ \citenamefont {Wrachtrup}}]{Schmid-Lorch2015Relaxometry}%
  \BibitemOpen
  \bibfield  {author} {\bibinfo {author} {\bibfnamefont {D.}~\bibnamefont {Schmid-Lorch}}, \bibinfo {author} {\bibfnamefont {T.}~\bibnamefont {Häberle}}, \bibinfo {author} {\bibfnamefont {F.}~\bibnamefont {Reinhard}}, \bibinfo {author} {\bibfnamefont {A.}~\bibnamefont {Zappe}}, \bibinfo {author} {\bibfnamefont {M.}~\bibnamefont {Slota}}, \bibinfo {author} {\bibfnamefont {L.}~\bibnamefont {Bogani}}, \bibinfo {author} {\bibfnamefont {A.}~\bibnamefont {Finkler}},\ and\ \bibinfo {author} {\bibfnamefont {J.}~\bibnamefont {Wrachtrup}},\ }\bibfield  {title} {\bibinfo {title} {Relaxometry and dephasing imaging of superparamagnetic magnetite nanoparticles using a single qubit},\ }\href {https://doi.org/10.1021/acs.nanolett.5b00679} {\bibfield  {journal} {\bibinfo  {journal} {Nano Letters}\ }\textbf {\bibinfo {volume} {15}},\ \bibinfo {pages} {4942} (\bibinfo {year} {2015})},\ \bibinfo {note} {pMID: 26218205},\ \Eprint {https://arxiv.org/abs/https://doi.org/10.1021/acs.nanolett.5b00679}
  {https://doi.org/10.1021/acs.nanolett.5b00679} \BibitemShut {NoStop}%
\bibitem [{\citenamefont {Barry}\ \emph {et~al.}(2020)\citenamefont {Barry}, \citenamefont {Schloss}, \citenamefont {Bauch}, \citenamefont {Turner}, \citenamefont {Hart}, \citenamefont {Pham},\ and\ \citenamefont {Walsworth}}]{Barry2020Sensitivity}%
  \BibitemOpen
  \bibfield  {author} {\bibinfo {author} {\bibfnamefont {J.~F.}\ \bibnamefont {Barry}}, \bibinfo {author} {\bibfnamefont {J.~M.}\ \bibnamefont {Schloss}}, \bibinfo {author} {\bibfnamefont {E.}~\bibnamefont {Bauch}}, \bibinfo {author} {\bibfnamefont {M.~J.}\ \bibnamefont {Turner}}, \bibinfo {author} {\bibfnamefont {C.~A.}\ \bibnamefont {Hart}}, \bibinfo {author} {\bibfnamefont {L.~M.}\ \bibnamefont {Pham}},\ and\ \bibinfo {author} {\bibfnamefont {R.~L.}\ \bibnamefont {Walsworth}},\ }\bibfield  {title} {\bibinfo {title} {Sensitivity optimization for nv-diamond magnetometry},\ }\href {https://doi.org/10.1103/RevModPhys.92.015004} {\bibfield  {journal} {\bibinfo  {journal} {Rev. Mod. Phys.}\ }\textbf {\bibinfo {volume} {92}},\ \bibinfo {pages} {015004} (\bibinfo {year} {2020})}\BibitemShut {NoStop}%
\end{thebibliography}%

\end{document}